\documentclass[twocolumn,twocolappendix]{aastex63}
\usepackage{ragged2e}
\usepackage{natbib}

\accepted{ApJ}

\shorttitle{Molecular Gas in Quasar Hosts}
\shortauthors{Molina, J. et al.}

\begin{document}

\title{Compact Molecular Gas Distribution in Quasar Host Galaxies}

\correspondingauthor{Juan Molina}
\email{jumolina@pku.edu.cn}

\author[0000-0002-8136-8127]{Juan Molina}
\affil{Kavli Institute for Astronomy and Astrophysics, Peking University, Beijing 100871, China}

\author[0000-0003-4956-5742]{Ran Wang}
\affil{Kavli Institute for Astronomy and Astrophysics, Peking University, Beijing 100871, China}
\affiliation{Department of Astronomy, School of Physics, Peking University, Beijing 100871, China}

\author[0000-0002-4569-9009]{Jinyi Shangguan}
\affil{Max-Planck-Institut f\"{u}r Extraterrestrische Physik (MPE), Giessenbachstr., D-85748 Garching, Germany}

\author[0000-0001-6947-5846]{Luis C. Ho}
\affil{Kavli Institute for Astronomy and Astrophysics, Peking University, Beijing 100871, China}
\affiliation{Department of Astronomy, School of Physics, Peking University, Beijing 100871, China}

\author[0000-0002-8686-8737]{Franz E. Bauer}
\affil{Instituto de Astrof{\'{\i}}sica and Centro de Astroingenier{\'{\i}}a, Facultad de F{\'{i}}sica, Pontificia Universidad Cat{\'{o}}lica de Chile, Casilla 306, Santiago 22, Chile}
\affiliation{Millennium Institute of Astrophysics (MAS), Nuncio Monse{\~{n}}or S{\'{o}}tero Sanz 100, Providencia, Santiago, Chile} \affiliation{Space Science Institute, 4750 Walnut Street, Suite 205, Boulder, Colorado 80301}

\author[0000-0001-7568-6412]{Ezequiel Treister}
\affil{Instituto de Astrof{\'{\i}}sica and Centro de Astroingenier{\'{\i}}a, Facultad de F{\'{i}}sica, Pontificia Universidad Cat{\'{o}}lica de Chile, Casilla 306, Santiago 22, Chile}

\author{Yali Shao}
\affil{Max-Planck-Institut f\"ur Radioastronomie, Auf dem H\"{u}gel 69, 53121 Bonn, Germany}
 
\begin{abstract}
We use Atacama Large Millimeter/submillimeter Array CO(2--1) observations of six low-redshift Palomar-Green quasars to study the distribution and kinematics of the molecular gas of their host galaxies at kpc-scale resolution. While the molecular gas content, molecular gas fraction, and star formation rates are similar to those of nearby massive, star-forming galaxies, the quasar host galaxies possess exceptionally compact, disky molecular gas distributions with a median half-light radius of 1.8\,kpc and molecular gas mass surface densities $\gtrsim 22 $\,$M_\odot$\,pc$^{-2}$.  While the overall velocity field of the molecular gas is dominated by regular rotation out to large radii, with rotation velocity-to-velocity dispersion ratio $\gtrsim 9$, the nuclear region displays substantial kinematic complexity associated with small-scale substructure in the gas distribution.  A tilted-ring analysis reveals that the kinematic and photometric position angles are misaligned on average by $\sim 34 \pm 26^\circ$, and provides evidence of kinematic twisting. These observations provide tantalizing clues to the detailed physical conditions of the circumnuclear environments of actively accreting supermassive black holes.
\end{abstract}


\section{Introduction}

Correlations between host galaxy bulge properties and supermassive black hole (BH) mass \citep{Kormendy1995,Ferrarese2000,Gebhardt2000,Tremaine2002} have led to the conclusion that galaxies and BHs coevolve \citep{Richstone1998,Kormendy2013}, perhaps mediated by feedback mechanisms from active galactic nuclei (AGNs) that regulate gas accretion toward the center and star formation on galactic scales (e.g., \citealt{Fabian2012,Heckman2014}).  AGN feedback can produce strong multi-phase gas outflows (e.g., \citealt{Cicone2014,Perna2015,Feruglio2015,Karouzos2016,Morganti2017,Cicone2018,Fluetsch2019}) that are powerful enough to heat or remove the interstellar medium from the host galaxy \citep{Silk1998,Harrison2018}, thereby suppressing subsequent star formation activity (e.g., \citealt{Dubois2016}) and keeping the host galaxy quiescent \citep{Fabian2012}. It is a key ingredient in numerical, theoretical and semi-analytic models to reproduce the lack of massive galaxies in the high-mass end of the mass function (e.g. \citealt{Kauffmann2000,Croton2006,Schaye2015,Sijacki2015,Lacey2016}).

Notwithstanding these compelling arguments, serious doubts remain as to whether AGN feedback effectively removes sufficient cold gas from the host galaxy to curtail its ongoing star formation activity. The interstellar medium content of the host galaxies of nearby AGNs shows no evidence of depletion relative to star-forming galaxies of similar stellar mass, based on gas masses derived from direct observations of neutral atomic hydrogen \citep{Ho2008,Fabello2011,Gereb2015,Zhu2015,Ellison2019AGN} and CO \citep{Maiolino1997,Evans2001,Scoville2003,Evans2006,Bertram2007,Shangguan2020,Jarvis2020}, as well as inferred indirectly from dust emission \citep{Shangguan2018,Shangguan2019} or dust extinction \citep{Zhuang2020,Yesuf2020}.  Far from being quenched, stars seem to form with even greater efficiency in the host galaxies of luminous AGNs \citep{Shangguan2020b,Zhuang2020}.  It should be further noted that AGN host galaxies possess not only a ``normal'' gas reservoir, but the gas appears largely kinematically regular, as evidenced by their global line widths in H~I \citep{Ho2008} and CO \citep{Shangguan2020b}, which are consistent with rotational support, and by the absence of significant molecular outflows \citep{Shangguan2020}.

The above-mentioned observations suggest that AGN feedback, if present, imparts only a modest, likely localized effect on the cold gas. For example, the existing molecular gas observations cannot rule out that the cold gas is in the process of being expelled from nuclei but still associated with the host galaxy.  And while AGNs can heat the molecular gas and suppress star formation through ``negative'' feedback \citep{Papadopoulos2010}, they can also exert the opposite effect---``positive'' feedback---that can compress the cold gas and enhance star formation \citep{Cresci2015,Carniani2016,Maiolino2017,Cresci2018,Gallagher2019}. From an observational point of view, the impact of AGN feedback is still far from settled, and it is essential to obtain spatially resolved information on the molecular gas in active galaxies to gain further insight into the physical processes that govern the coevolution of supermassive BHs and their host galaxies.

Quasars, the most luminous of the active galaxies, are the ideal sites to investigate the possible interplay between AGN feedback and the molecular gas of their host galaxies.  In the popular merger-driven evolutionary scenario of quasars \citep{Sanders1988}, two gas-rich galaxies merge, gravitational torques drive the cold gas to the center of the merger remnant, and vigorous starburst activity and BH growth ensue. The prodigious energy released by the AGN expels the enshrouding gas and dust, giving birth to an optically visible, largely unobscured quasar \citep{Hopkins2008}. 

The frequent association of a young stellar population with quasar host galaxies supports the notion that star formation accompanies or precedes AGN activity (e.g., \citealt{Canalizo2001,Canalizo2013,Jahnke2007,Kim2019,Zhao2019}), in qualitative agreement with the merger-induced evolution scenario of quasars.

However, stellar morphology studies of quasar host galaxies suggest that only a fraction of the systems show tidal and/or dynamical perturbation signatures that may be associated to recent merger activity (e.g. \citealt{Veilleux2009}). Not all the quasar hosts with enhanced star formation activity show evidence of dynamical perturbations \citep{Shangguan2020b}. The equal mass merger scenario may be applicable to ultraluminous infrared galaxies and AGNs hosted in ellipticals. But, in other cases, unequal mass mergers which do not perturb the more massive interacting galaxy may be required.

\begin{figure}
\centering
\includegraphics[width=0.9\columnwidth]{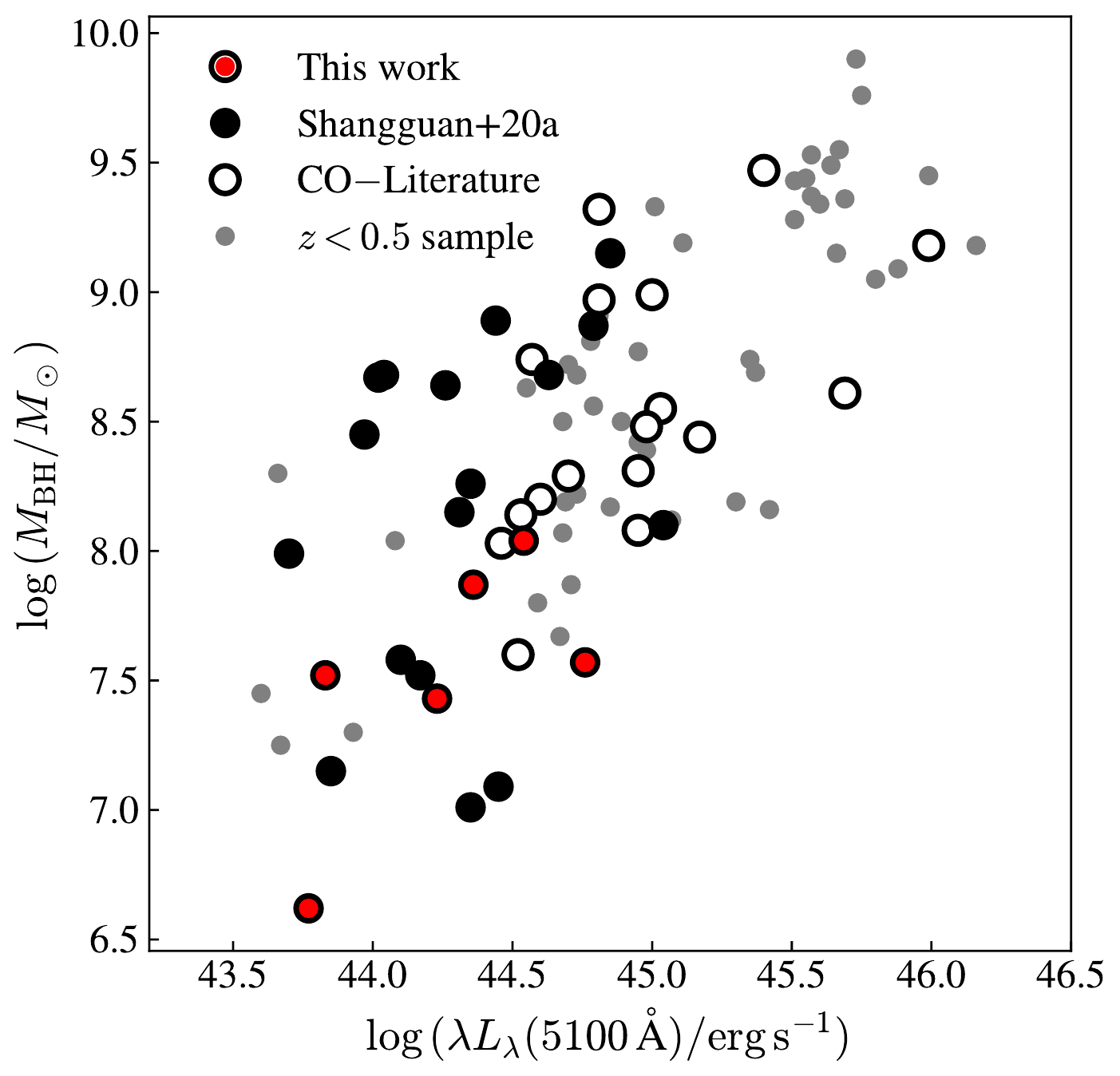}
\caption{\label{fig:PG_sample} Black hole mass as a function of the AGN monochromatic luminosity at 5100\,\r{A} for the PG quasar host galaxies. We show separately the host galaxies with CO data from literature (open circles), our ACA observations (filled circles; \citealt{Shangguan2020}) and the six targets observed by ALMA (red-filled circles). The six targets sample the $M_{\rm BH} \lesssim 10^8$\,$M_\odot$ limit within the PG survey.}
\end{figure}

This work reports new, relatively high-resolution (beam $\sim 0\farcs4-1\arcsec$, which corresponds to physical scales of $\lesssim 1$ kpc) Atacama Large Millimeter/sub-millimeter array (ALMA) molecular gas observations toward the host galaxies of six Palomar-Green (PG) quasars \citep{Boroson1992}. The PG survey contains 87 optically/UV-selected low-redshift ($z < 0.5$) type~1 (broad-lined) quasars \citep{Boroson1992}. This sample has been studied extensively, enjoying a rich repository of multi-wavelength data for the AGN and host galaxy, including optical spectra \citep{Boroson1992,Ho2009}, radio properties \citep{Kellermann1989,Kellermann1994}, X-ray constraints \citep{Reeves2000,Bianchi2009}, dust properties for both the torus and host galaxy \citep{Shi2014,Petric2015,Shangguan2018,Zhuang2018}, high-resolution optical and near-infrared imaging of the stellar component of the host galaxies \citep{Kim2008,Kim2017,Zhang2016,Zhao2020}, and star formation rates (SFRs, \citealt{Xie2020}).  

We benefit from our Cycle~5 Atacama Compact (Morita) Array (ACA) survey \citep{Shangguan2020}, which targeted the carbon monoxide molecule ($^{12}$CO) $J=2 \rightarrow 1$ transition [$\nu_{\rm rest} = 230.538$\,GHz; hereinafter CO(2--1)] for a subset of 23 $z < 0.1$ PG quasars selected from our previous infrared study \citep{Shangguan2018}. Along with other molecular gas observations reported in the literature, we now have CO measurements for a representative subset of 40 PG quasars at $z < 0.3$ \citep{Shangguan2020}.  The main goal of this study is to characterize in greater detail the spatial distribution and kinematics of the molecular gas for the small subset of PG quasars with brightest CO emission. By using these selection criteria, we aim to maximize source detection coverage and facilitate the morpho-kinematic analysis. We employed no further constraint to select our targets.

\begin{table*}
	\centering
	\def\arraystretch{1.5}
	\setlength\tabcolsep{4pt}
    	\caption{\label{tab:hostprop} Basic Parameters of the Sample}
    	\vspace{1mm}
	\begin{tabular}{cccccccccccc} 
		\hline
		\hline
		Object & R.A. & Decl. & $z$ & $D_L$ & Morph. & $\log M_\star$ & $\log L_{\rm IR}$ & SFR & $\log M_{\rm gas}$ & $\log M_{\rm BH}$ & $\log \lambda L_\lambda(5100$\r{A}$)$ \\
		& (J2000.0) & (J2000.0) & & (Mpc) & &($M_\odot$) & (erg\,s$^{-1}$) & ($M_\odot$\,yr$^{-1}$) & ($M_\odot$) & ($M_\odot$) & (erg\,s$^{-1}$)\\
		(1) & (2) & (3) & (4) &(5) & (6) & (7) & (8)& (9) & (10) & (11) & (12) \\
		\hline
		PG\,0050+124    & 00:53:34.94 & +12:41:36.2    & 0.061 & 282.3 & Disk & 11.12 & 44.94$^{+0.01}_{-0.01}$ & 26.3$^{+0.7}_{-0.5}$ & 10.3 & 7.57 & 44.76\\
         PG\,0923+129    & 09:26:03.29 & +12:44:03.6    & 0.029 & 131.2 & Disk & 10.71 & 44.05$^{+0.01}_{-0.02}$ & 3.4$^{+0.1}_{-0.2}$ & 9.5 & 7.52 & 43.83\\
         PG\,1011$-$040 & 10:14:20.69 & $-$04:18:40.5 & 0.058 & 267.9 & Disk & 10.87 & 43.98$^{+0.02}_{-0.02}$ & 2.9$^{+0.2}_{-0.2}$ & 9.7 & 7.43 & 44.23\\
         PG\,1126$-$041 & 11:29:16.66 & $-$04:24:07.6 & 0.060 & 277.5 & Disk & 10.85 & 44.46$^{+0.03}_{-0.03}$ & 8.7$^{+0.7}_{-0.6}$ & 9.7 & 7.87 & 44.36\\
         PG\,1244+026    & 12:46:35.25 & +02:22:08.8    & 0.048 & 220.1 & Disk & 10.19 & 43.85$^{+0.02}_{-0.01}$ & 2.1$^{+0.1}_{-0.1}$ & 8.8 & 6.62 & 43.77\\
         PG\,2130+099    & 21:32:27.81 & +10:08:19.5    & 0.061 & 292.3 & Disk & 10.85 & 44.37$^{+0.02}_{-0.03}$ & 7.1$^{+0.4}_{-0.5}$ & 9.7 & 8.04 & 44.54\\
		\hline                                                                             
	\end{tabular}
	\justify
	{\justify \textsc{Note}--- (1) Source name. (2) Right ascension. (3) Declination. (4) Redshift. (5) Luminosity distance. (6) Host galaxy morphology type based on HST image and taken from \citep{Zhang2016,Kim2017,Yulin2020} (7) Stellar mass; the $1\,\sigma$ uncertainty is 0.3\,dex \citep{Shangguan2018}. (8) Total infrared luminosity of the host galaxy \citep{Shangguan2018}. (9) Star formation rate derived from IR luminosity by adopting Eq.~4 of \citep{Kennicutt1998b} and a Kroupa Initial mass function \citep{Kroupa2001}. (10) Total gas mass inferred from dust mass measurements. The $1\,\sigma$ uncertainty is 0.2\,dex \citep{Shangguan2018}. (11) Black hole mass, estimated by applying the calibration of \citet{Ho2015} using the AGN monochromatic luminosity at 5100\,\r{A} (Col. 12) and the H$\beta$ line width \citep{Shangguan2018}. (12) AGN monochromatic luminosity at 5100\,\r{A}.}	
	
\end{table*}

\begin{table*}
        \centering
        \def\arraystretch{1.0}
        \setlength\tabcolsep{4pt}
        \caption{\label{tab:obs} ALMA Cycle~6 Observational Setup}
        \vspace{1mm}
        \begin{tabular}{cccccccccc}
                \hline
                \hline
                Object & Observation & Bandpass \& & Phase & PWV & On-source & Beam Size & Beam Position& RMS \\
                 &  Date & Flux Calibrator & Calibrator & (mm) & Time (min) & ($\arcsec \times\arcsec $) & Angle ($^\circ$) &(mJy\,beam$^{-1}$) \\
                (1) & (2) & (3) & (4) & (5) & (6) & (7) & (8) & (9) \\
                \hline
         PG\,0050+124    & 26\,Nov.\,2018 & J0006$-$0623 & J0121+1149    & 1.49 & 11.65 & $0.40 \times 0.36$ & 21.1 &  0.55\\
         PG\,0923+129    & 17\,Mar.\,2019 & J1058+0133    & J0853+0654    & 1.57 & 16.72 & $1.27 \times 1.00$ & 71.6 & 0.47\\
         PG\,1011$-$040 & 14\,Mar.\,2019 & J1037$-$2934 & J1010$-$0200 & 2.07 & 17.22 & $1.41 \times 0.91$ & 72.0 & 0.59\\
         PG\,1126$-$041 & 14\,Mar.\,2019 & J1058+0133    & J1131$-$0500 & 1.41 & 36.95 & $1.25 \times 0.89$ & 83.1 & 0.37\\
         PG\,1244+026    & 19\,Mar.\,2019 & J1256$-$0547 & J1239+0730    & 1.62 & 31.40 & $1.41 \times 1.17$ & 95.4 & 0.41\\
         PG\,2130+099    & 22\,Mar.\,2019 & J2000$-$1748 & J2147+0929    & 0.96 & 40.52 & $1.31 \times 1.13$ & 25.5 & 0.34\\
                \hline
        \end{tabular}
\end{table*}

The paper is organized as follows. Our sample, measurements, and comparison sample are described in Section~\ref{sec:obs}. Section~\ref{sec:Analysis_and_results} presents the models used to analyze the data and their basic results. We further discuss our findings from a global perspective in Section~\ref{sec:Discussion}, and we conclude in Section~\ref{sec:Conclusions}. We adopt a $\Lambda$CDM cosmology with $\Omega_m = 0.308$, $\Omega_\Lambda = 0.692$, and $H_0 = 67.8$\,km\,s$^{-1}$\,Mpc$^{-1}$ \citep{Planck2016}.

\section{Sample and Observations}
\label{sec:obs}

\subsection{ALMA Observations and Data Reduction}
\label{Obs_setup}

We analyze Cycle~6 ALMA follow-up observations (program 2018.1.00006.S; PI: F. Bauer) of the six PG quasar host galaxies (Table~\ref{tab:hostprop}) taken from our $z < 0.1$ PG quasar subsample \citep{Shangguan2020}. These six objects constitute the brightest CO(2--1) sources within the ACA sample, which enables us to obtain deep, $\lesssim 0\farcs4$--$1\farcs4$-scale imaging with reasonable exposure times. They are representative of low-$z$ PG quasars with low to moderate BH masses and luminosities (Figure~\ref{fig:PG_sample}) and disk-like stellar morphology (Table~\ref{tab:hostprop}).  The ALMA observations were carried during November 2018 to March 2019 in good weather conditions with precipitate water vapor (PWV) $\lesssim 2.1$\,mm. All the targets were observed by 43 antennas. These observations were designed to detect the continuum and CO(2--1) line in Band~6 using four spectral windows, each covering 1.875\,GHz in bandwidth with a spectral resolution of 7.8125\,MHz, equivalent to a channel resolution of $\sim 11$\,km\,s$^{-1}$. The observational setup for each source is described in Table~\ref{tab:obs}. The ALMA flux calibration uncertainty is $\lesssim 10$\,\% \citep{Fomalont2014,Bonato2018}.

We use the Common Astronomy Software Application (\textsc{CASA}; \citealt{McMullin2007}) to reduce the ALMA data, employing the standard pipeline to calibrate the data to generate the $uv$ visibilities.  To minimize missing flux from possible extended source emission and to obtain more sensitive imaging of the total CO(2--1) emission, we concatenate our Cycle~6 observations with the previous Cycle~5 ACA observations using the task \textsc{concat}. For the ACA data we consider the same spectral channel flagging that \citet{Shangguan2020} employed in their work, while for the 12-m antenna data we flag the spectral channels near a sky feature found at $\sim 233.7$\,GHz. The flagged channels are then input to the task \textsc{uvcontsub} to subtract the continuum, before proceeding with imaging the line emission. 

The emission-line imaging is performed using \textsc{tclean} with robust weighting (\textsc{robust} = 0.5) and a channel resolution of $11$\,km\,s$^{-1}$. The spatial pixel scale is set to sample the synthesized beam by five pixels.  Visual inspection of the masks constructed by the masking procedure \textsc{auto-multithresh} indicates that optimal results can be obtained by setting the noise, sidelobe, and low-noise thresholds to 4.0, 1.0, and 1.5, respectively, and the fractional beam size to 0.4. All other parameters of \textsc{auto-multithresh} were left at their default values, as recommended for combined ACA and 12-m data.\footnote{https://casaguides.nrao.edu/index.php?title=Automasking\_Guide} The RMS values and the sizes of the synthesized beam are given in Table~\ref{tab:obs}. 

\begin{figure}
\centering
\includegraphics[width=1.0\columnwidth]{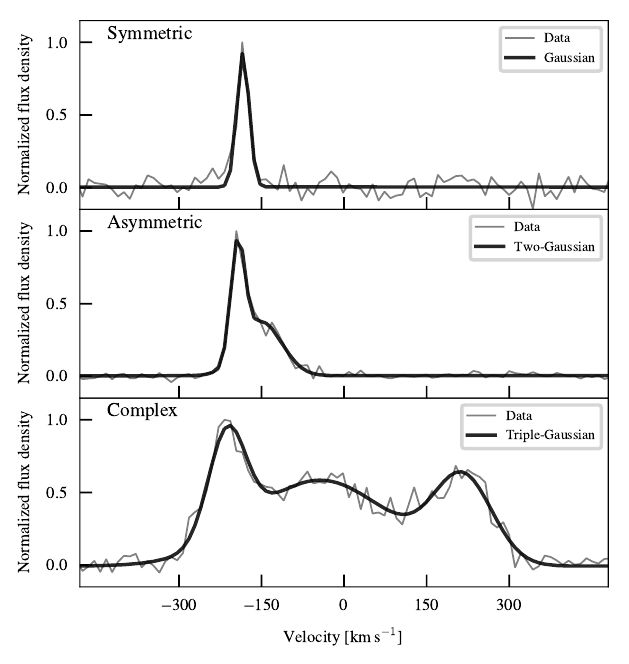}
\caption{\label{fig:emline_shapes} Example of the three types of CO(2--1) line profiles observed for PG\,1126$-$041. These are representative of the emission-line shapes seen across the six targets.}
\end{figure}

\begin{figure*}
\centering
\includegraphics[width=2.0\columnwidth]{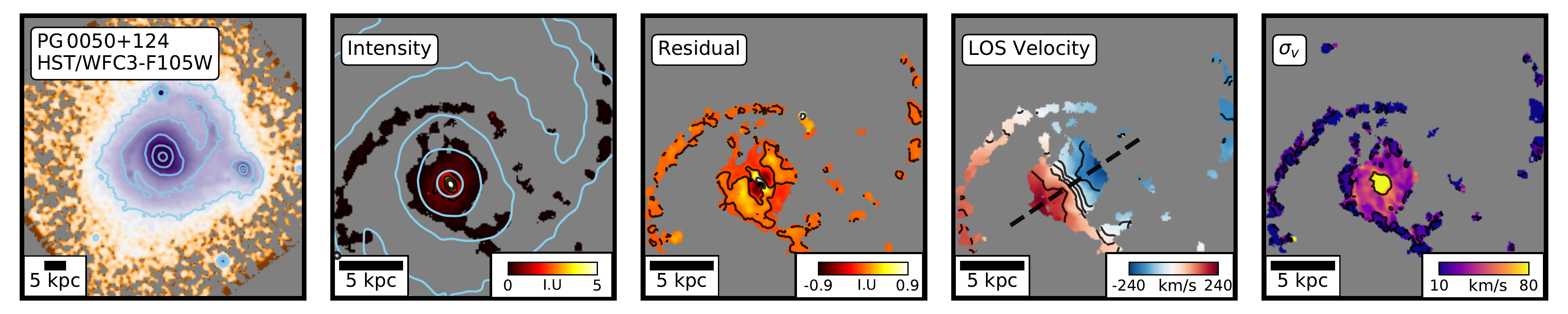}\\
\includegraphics[width=2.0\columnwidth]{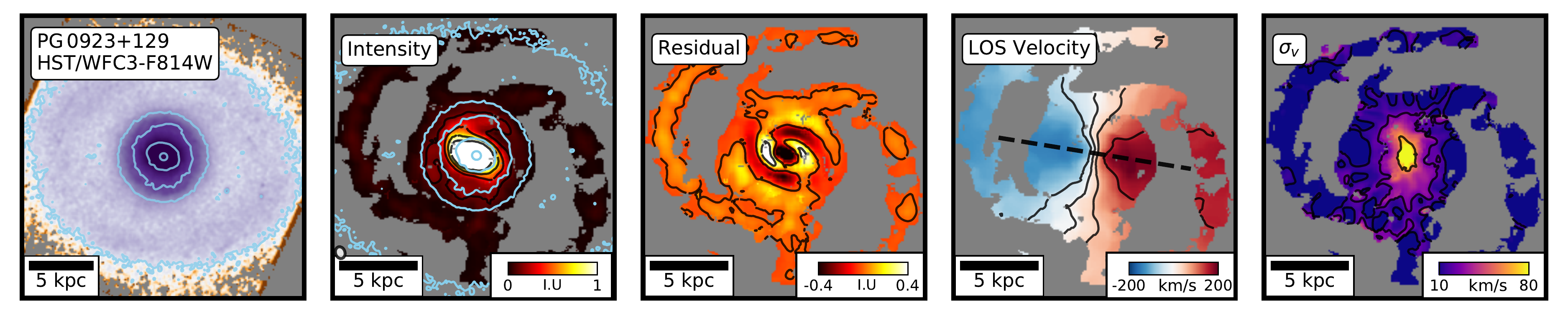}\\
\includegraphics[width=2.0\columnwidth]{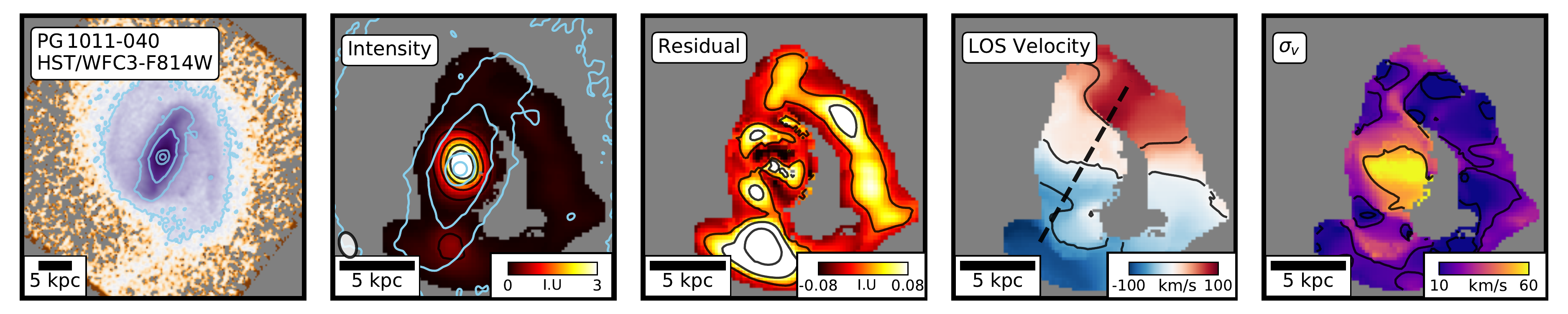}\\
\includegraphics[width=2.0\columnwidth]{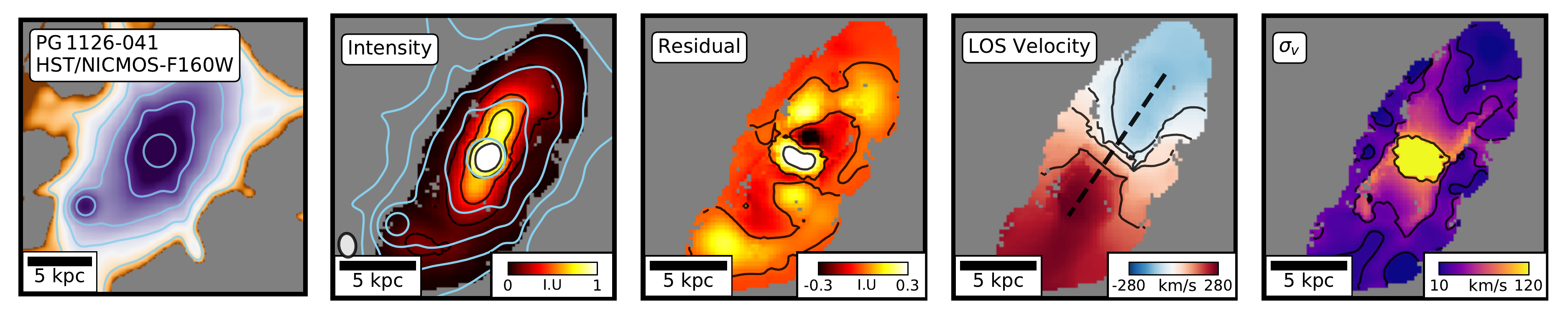}\\
\includegraphics[width=2.0\columnwidth]{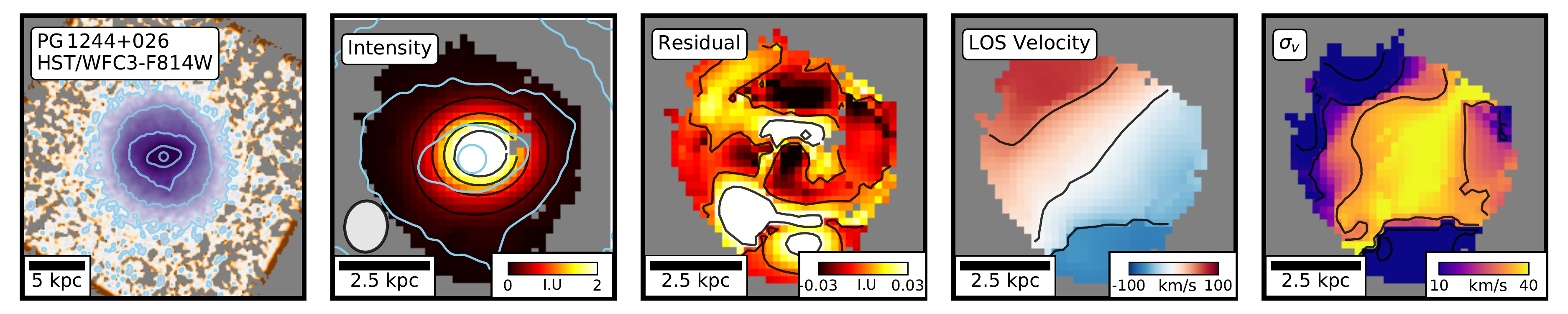}\\
\includegraphics[width=2.0\columnwidth]{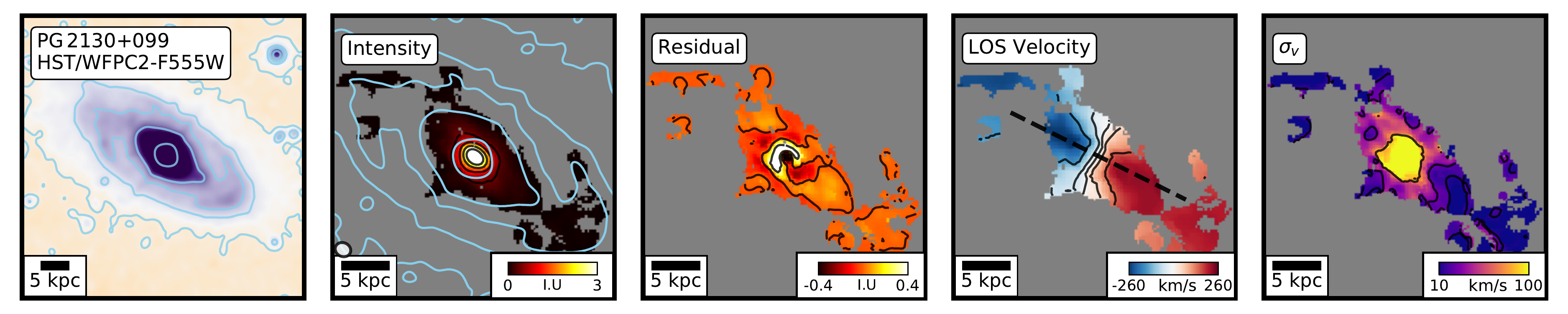}\\
\caption{\label{fig:int_maps} Hubble Space Telescope (HST) image taken from \citet{Yulin2020}, CO(2--1) line intensity, best-fit intensity residuals (model fits described in Section 3.2), LOS velocity, velocity dispersion, and map of emission-line shapes for the PG quasars.  In the first and second columns, the light blue contours are taken from each HST image and correspond to the $[0.2, 1, 3, 10, 100] \times 10^{-3}$ levels relative to the maximum brightness value (for PG\,2130+099 we show the $[3, 3.5, 4, 5, 20] \times 10^{-2}$ relative levels to improve visualization). The HST images reveal the stellar component emission of quasar host galaxies. The beam is plotted above the scale bar of the CO(2--1) intensity map. In the LOS velocity map, the dashed line represents the direction of the kinematic major axis.}
\end{figure*}

\subsection{Emission-line Characterization and Map Construction}
\label{sec:em-lines}

We implement a three-step procedure to characterize the CO(2--1) emission-line shapes.  For each pixel, we first average the spectrum by considering a box/squared region with size comparable to the synthesized beam (e.g., \citealt{Swinbank2012a}), and we estimate the noise from the line-free channels. We fit a Gaussian model to the spectrum using the least-squares minimization procedure implemented in the \textsc{Python} package \textsc{lmfit} \citep{Newville2014}.  As initial guesses, we assume that the line centroid equals to the line peak location in the spectrum, and that the line width equals to 20\,km\,s$^{-1}$. The line width is restricted to a maximum value of 500\,km\,s$^{-1}$.  We use the Bayesian information criterion (BIC; \citealt{Schwarz1978}), which penalizes by the model parameter number, instead of a $\chi^2$-based criterion to determine whether a line is detected.  We estimate the likelihood of the best-fit Gaussian model by comparing it with a straight line fit (i.e., no emission line present). We consider a $5\,\sigma$ probability threshold for detection. If this threshold is not achieved, we bin the spectrum over a larger area by increasing the size of the extraction box by one pixel per side, and then repeat the fit until either the probability threshold is achieved or the third iteration is reached. If no detection is achieved after three iterations, we assume that no emission is present, mask the pixel, and skip to the next one. We stop at the third binning iteration in order to avoid large binned zones when compared to the beam size. We also exclude highly uncertain models by masking pixels that have lines with peak signal-to-noise (S/N) lower than 3.  Pixels with high S/N often show asymmetric, occasionally highly complicated line shapes.  When an emission line is detected, we increase the number of Gaussian components and we repeat the fit. We compute the multi-Gaussian model likelihood by calculating BIC with respect to the last accepted model. Again, we consider a $5\,\sigma$ probability threshold for model acceptance. Most spectra that are symmetric can be well fit with a single Gaussian component, those that are asymmetric can be described by two Gaussians, and three components are necessary for even more complex shapes.  Figure~\ref{fig:emline_shapes} gives some examples for the case of PG\,1126$-$041.  We find that multiple Gaussians perform better than a high-order Gauss-Hermite function \citep{vanderMarel1993}.  

The least-squares minimization technique can be highly sensitive to the given initial guesses.  To mitigate this issue, we refit the detected emission lines by considering a series of new initial guesses taken from the best-fit parameters obtained from the neighboring pixels and from the pixel itself during the first step.  The neighbor pixels are defined as those within the binning box used to extract the averaged spectrum. We select the final model that gives the lowest BIC.  Note that the fitting procedure can give rise to false positive detections from noisy peaks in the spectrum.  We mask these noisy pixels by applying a procedure that mimics the pruning routine employed by the \textsc{auto-multithresh} task within \textsc{tclean}, by masking detected pixel groups that have projected sizes smaller than 0.6 times the beam size. 

Finally, we use Monte Carlo resampling to derive the model parameter uncertainties. We measure the average spectrum noise level for each pixel and assume that the noise follows a normal distribution. We then add the simulated noise to the observed spectrum and fit the line. We iterate 300 times to obtain a distribution for each parameter, and we estimate the $1\,\sigma$ uncertainties of the parameter from the 16th and 84th percentiles of the distribution.

The intensity, line-of-sight (LOS) velocity, and velocity dispersion maps are shown in Figure~\ref{fig:int_maps}. The latter two maps are constructed by calculating the luminosity-weighted moment one and two from the best-fit models of the emission line in each pixel.

We perform a sanity check to determine how much CO(2--1) emission is missed by our line shape fitting procedure.  We calculate the total velocity-integrated line intensity ($S_{\rm CO(2-1)} \Delta v$) from the intensity maps and we compare these to the values obtained by summing the flux densities from the channels above the 2\,$\sigma$ level for each target.  We find that our emission-line fitting procedure recovers $\sim 97$\,\% of  $S_{\rm CO(2-1)} \Delta v$ for the two worst cases (PG\,0050+124, PG\,0923+129), while for the best case (PG\,1011$-$040) it estimates $\sim 7$\,\% more $S_{\rm CO(2-1)} \Delta v$ value. These estimates indicate that our emission line procedure recovers most of the reliable CO(2--1) emission recorded in the datacubes.

\subsection{CO(2--1) Luminosity and Molecular Gas Estimates}
\label{sec:COlum}

We estimate the $S_{\rm CO(2-1)} \Delta v$ values by summing all the pixel values from the intensity maps. The luminosity $L^\prime_{\rm CO(2-1)}$ is calculated following \citep{SV2005}

\begin{equation}
\label{eq:LCO}
L^\prime_{\rm CO(2-1)} = 3.25 \times 10^7\,\frac{S_{\rm CO(2-1)} \Delta v\, D^2_{L}}{\nu_{\rm obs}^{2}\,(1+z)^{3}}\,\,\, {\rm [K\,km\,s^{-1}\,pc^2]}, 
\end{equation}

\noindent where $S_{\rm CO(2-1)} \Delta v$ is in units of Jy\,km\,s$^{-1}$, $\nu_{\rm obs}$ is the observed frequency of the line in GHz, $D_{L}$ is the luminosity distance in Mpc, and $z$ is the redshift.  The CO(2--1) luminosities are used to estimate the CO(1--0) luminosity by adopting a luminosity ratio $L^\prime_{\rm CO(2-1)} / L^\prime_{\rm CO(1-0)} = 0.62$, the median value found by \cite{Shangguan2020} for eight PG quasar host galaxies. We obtain molecular gas masses assuming a CO-to-H$_2$ conversion factor $\alpha_{\rm CO} = 3.1$\,$M_\odot$\,(K\,km\,s$^{-1}$\,pc$^2$)$^{-1}$ with 0.3\,dex uncertainty \citep{Sandstrom2013}, a value consistent with dust-based gas masses independently derived for the PG quasars \citep{Shangguan2020}.

\begin{figure}
\centering
\includegraphics[width=1.0\columnwidth]{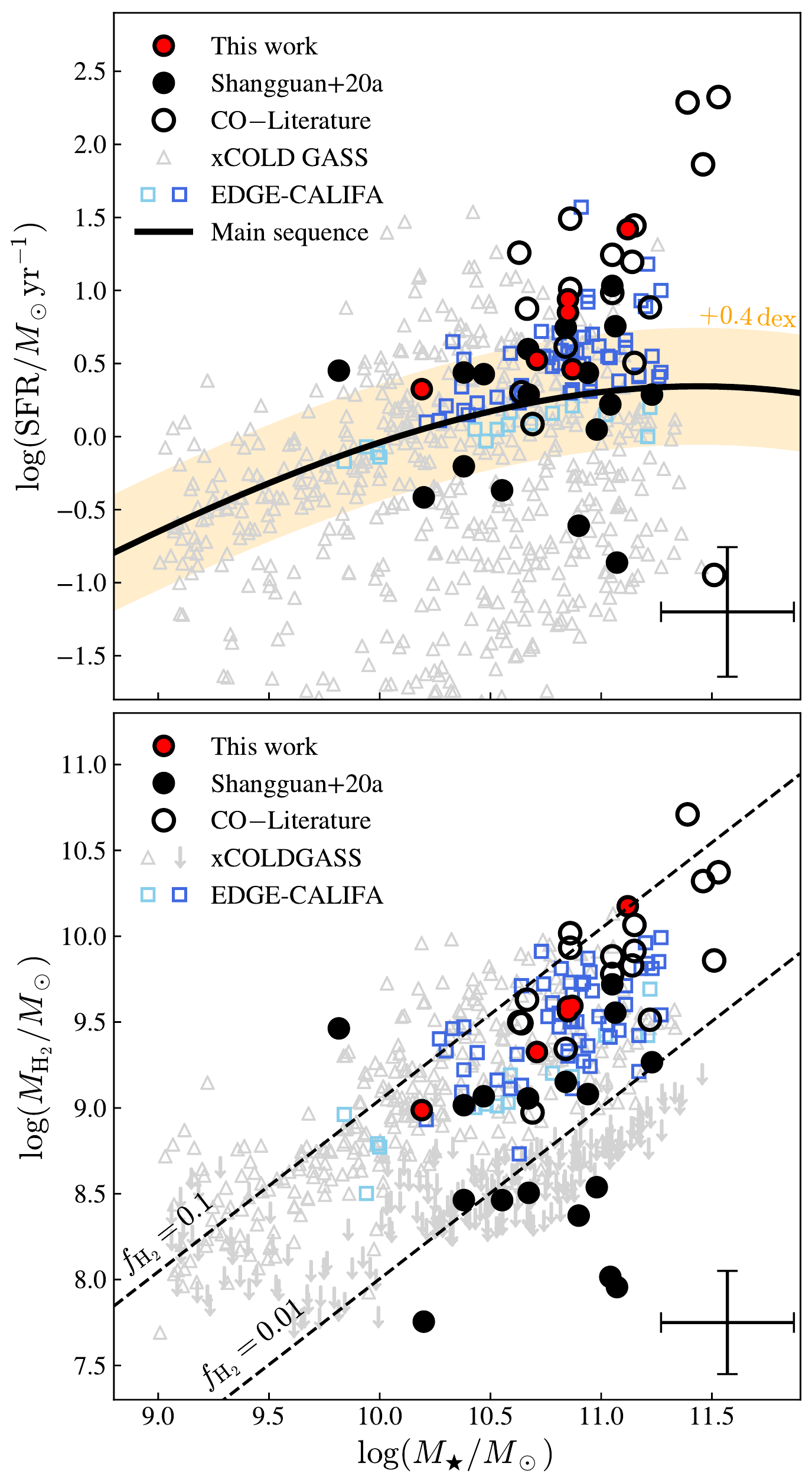}
\caption{\label{fig:comparison_sample} PG quasar host galaxies compared to the EDGE-CALIFA star-forming galaxies with measured CO sizes \citep{Bolatto2017}. (Top) The $M_\star$--SFR plane. We show host galaxies with CO data, color-coded following Figure~\ref{fig:PG_sample}. The EDGE-CALIFA data are presented in light blue and blue squares, the latter representing the subsample chosen to compare with our observations. The grey triangles show the xCOLD GASS sample \citep{Saintonge2017}. The solid line indicates the main sequence suggested by \citet{Saintonge2016}, with the orange shaded band indicating the $\pm 0.4$\,dex scatter of the main sequence. The typical uncertainty of the host galaxy measurements are represented by the error bar in the bottom-right corner. (Bottom) The $M_\star$--$M_{\rm H_2}$ plane. The data are color-coded as in the top panel, with the exception that upper limits for $M_{\rm H_2}$ for the xCOLD\,GASS galaxies are indicated by downward arrows. The dashed lines indicate values of constant $f_{\rm H_2}$. The PG quasar host galaxies and the EDGE-CALIFA survey overlap in both parameter spaces.}
\end{figure}

\subsection{Comparison Sample}
\label{sec:Csample}

Our analysis (Section~\ref{sec:comparison_res}) will compare the properties of quasar hosts with those of inactive galaxies.  We choose, for comparison, the EDGE-CALIFA sample \citep{Bolatto2017}, an interferometric CO(1--0) study with the Combined Array for Millimetre-wave Astronomy of 126 nearby (23--130\,Mpc) galaxies selected from the CALIFA survey \citep{Sanchez2012}. With a spectral resolution of $\sim 10$\,km\,s$^{-1}$ and an average spatial resolution of $\sim 1.4$\,kpc, the EDGE-CALIFA observations are well-matched to our ALMA observational setup. 

\citet{Bolatto2017} calculate molecular gas masses\footnote{\citet{Bolatto2017} adopt $\alpha_{\rm CO} = 0.8$\,$M_\odot$\,(K\,km\,s$^{-1}$\,pc$^2$)$^{-1}$ for the ultraluminous infrared galaxy Arp~220. We follow their convention.} assuming $\alpha_{\rm CO} = 4.36$\,$M_\odot$\,(K\,km\,s$^{-1}$\,pc$^2$)$^{-1}$, similar to but somewhat higher than our preferred value of $\alpha_{\rm CO} = 3.1$\,$M_\odot$\,(K\,km\,s$^{-1}$\,pc$^2$)$^{-1}$.  For consistency with our convention, we scale the EDGE-CALIFA molecular gas masses by a factor of $\sim 0.71$. We only consider the EDGE-CALIFA galaxies with measured CO sizes (69 systems). We further discard five sources that are likely to be AGN hosts based on the optical line intensity ratio diagnostics of \citet{Baldwin1981} and H$\alpha$ linewidth analysis \citep{Lacerda2020}. These five AGN hosts do not show any particular trend when compared to the main EDGE-CALIFA sample and only one of these (MRK\,79) is classified as type-I AGN ($\log \lambda L_\lambda(5100$\r{A}$) \sim 43.16$\,erg\,s$^{-1}$; \citealt{Lu2019}). Therefore, our comparison sample corresponds to a total of 64 galaxies taken from the EDGE-CALIFA survey.

Figure~\ref{fig:comparison_sample} compares the PG quasar host galaxies (the six sources mapped by ALMA plus the larger sample of sources with CO observations from \citealt{Shangguan2020}) with the EDGE-CALIFA galaxies, in terms of their SFR, stellar mass ($M_\star$), molecular gas mass ($M_{\rm H_2}$), and molecular gas mass fraction [$f_{\rm H_2} \equiv M_{\rm H_2}/(M_\star + M_{\rm H_2})$].  To improve further the statistics of the inactive galaxies, we also include the larger sample of nearby galaxies from xCOLD\,GASS \citep{Saintonge2017}.  As discussed in \citet[see also \citealt{Xie2020}]{Shangguan2020b}, the PG quasars generally track the main sequence of star-forming galaxies as defined by \citet{Saintonge2016}, with a non-negligible fraction lying significantly above it, to the extent that they can be deemed starburst systems.  The ALMA sample, by virtue of their selection, is biased toward higher SFRs, $M_{\rm H_2}$, and $f_{\rm H_2}$ compared to the overall PG sample with CO observations: all six objects lie on or above the main sequence, half of them formally exceeding the main sequence $1\,\sigma$ scatter of 0.4~dex.  The EDGE-CALIFA subsample overlaps well with the overall sample of quasar hosts in terms of $M_\star$, SFR, $M_{\rm H_2}$, and $f_{\rm H_2}$, but for the purposes of achieving a better match with the ALMA-mapped quasars, we further distinguish the EDGE-CALIFA galaxies that lie above the main sequence (blue squares in Figure~\ref{fig:comparison_sample}).

\begin{figure}
\centering
\includegraphics[width=0.9\columnwidth]{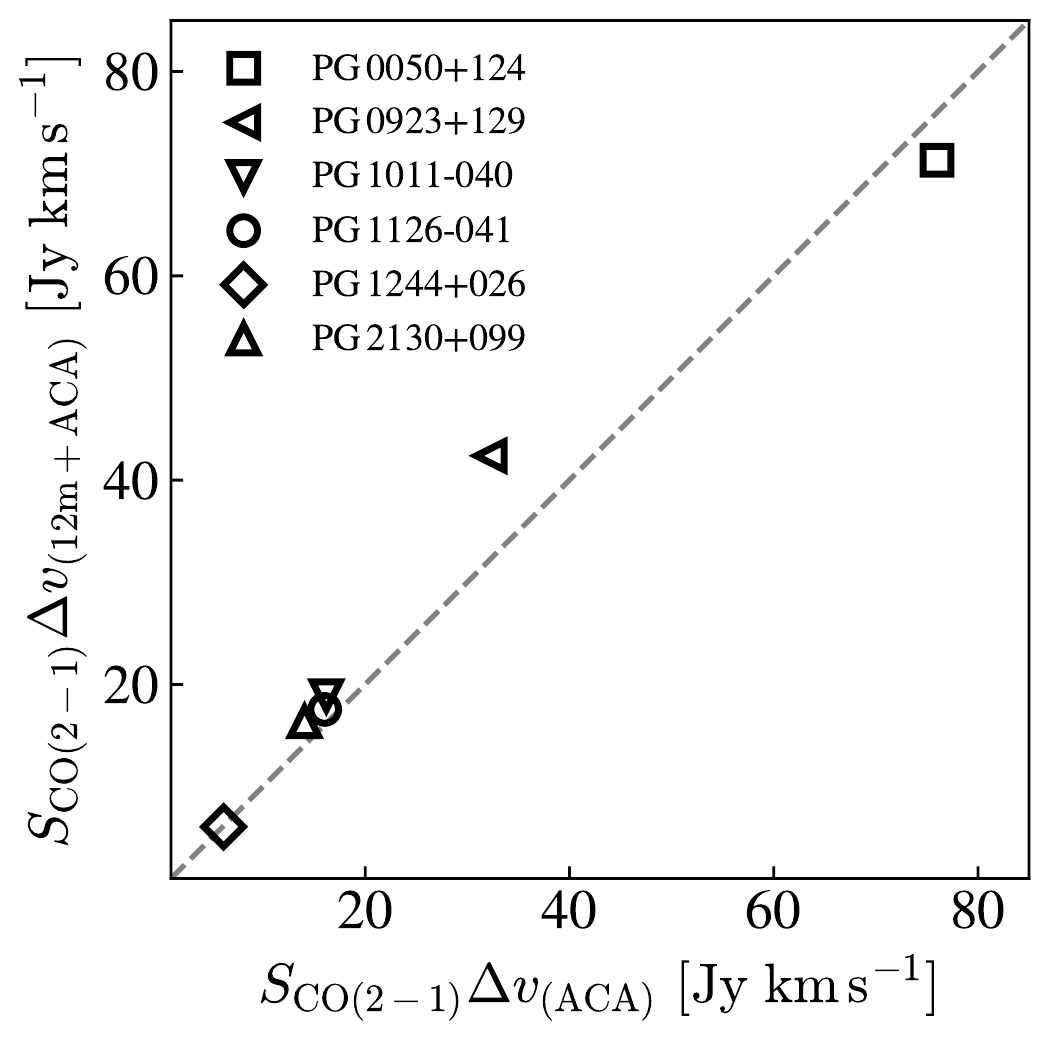}
\caption{\label{fig:int_fluxes} Comparison between the concatenated (12m + ACA) data and the ACA velocity-integrated CO(2--1) emission-line fluxes published in \citet{Shangguan2020}. The dashed line show the unity ratio. The $1\,\sigma$ errorbars of the data (without including flux calibration uncertainties) are smaller than the symbol sizes. We find good agreement between both estimates ($\Delta[S_{\rm CO(2-1)} \Delta v] \lesssim 24\,\%$), suggesting that the concatenated data can recover well both the compact and extended CO(2--1) emission.}
\end{figure}

\section{Analysis and Results}
\label{sec:Analysis_and_results}

\subsection{Sensitivity of the Concatenated Data Compared to the ACA Data}
\label{sec:int_flux}

To verify whether our process of data concatenation (12~m + ACA) properly recovers the CO line emission on extended scales, Figure~\ref{fig:int_fluxes} compares $S_{\rm CO(2-1)} \Delta v$ measured from our intensity maps with the ACA-only values reported by \citet{Shangguan2020}.  The two sets of measurements show good agreement, but there is a mild tendency for the concatenated data to recover somewhat higher fluxes than the ACA data alone.  Four of the six sources recover extra CO emission, ranging from $\sim 9\%$ to $24\%$, with an average of $\sim 12\%$.  This can be attributed to the higher sensitivity of the concatenated data compared to the ACA observations. For example, Figure~6 of \citet{Shangguan2020} shows that the CO emission of PG\,0923$+$129 is confined to the inner $\lesssim 8\arcsec$ region of the galaxy, whereas our map (Figure~\ref{fig:int_maps}) shows that the host galaxy exhibits CO spiral arms that extend up to $\sim 14\arcsec$ ($\sim$8.7\,kpc) from its center; the ACA observations miss the flux coming from these spiral arms.  The flux of PG\,2130$+$099 offers another example. Whereas the integrated spectrum based on ACA observations does not show an unambiguous double-horned profile characteristic of unresolved galactic rotation (Figure~6 of \citealt{Shangguan2020}), our concatenated data clearly do show disk-like rotation (Figure~\ref{fig:int_maps}), again indicating that the ACA data miss flux from the outskirts of this galaxy.  PG\,1244$+$026 is practically indistinguishable (within $\sim 1\%$) between the two data sets, perhaps to be expected considering its highly compact gas distribution.  Only PG\,0050$+$124 (I\,Zw\,1) stands with its $\sim 6\%$ flux deficit in the concatenated data set, but this is insignificant compared to the $\sim 5\%-10\%$ uncertainty of the absolute flux scale \citep{Fomalont2014,Bonato2018}.  We conclude that the concatenated data are sensitive enough to trace most of the CO(2--1) line emission coming from the central part of the host galaxies.

\begin{figure*}
\centering
\includegraphics[width=0.67\columnwidth]{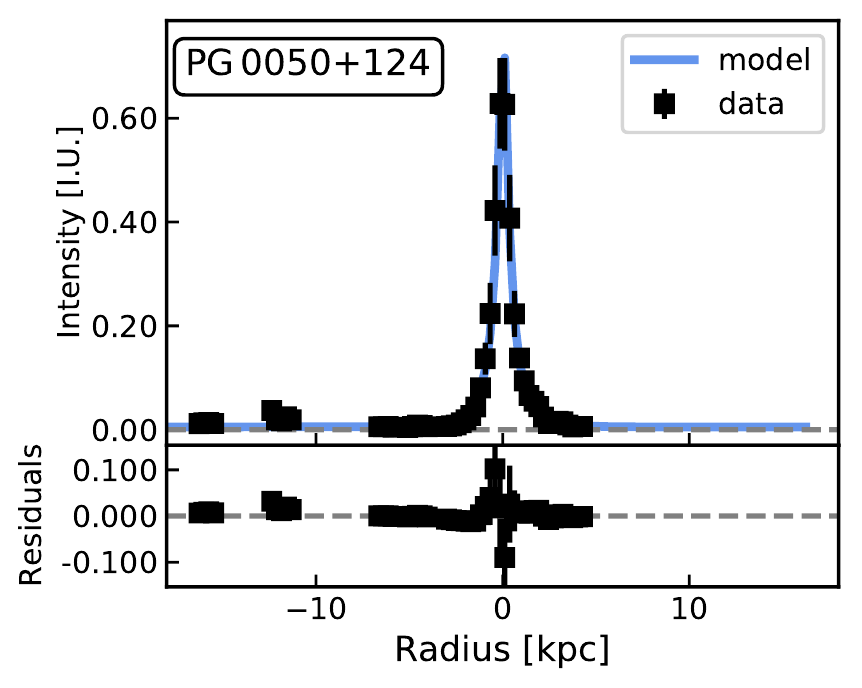}
\includegraphics[width=0.67\columnwidth]{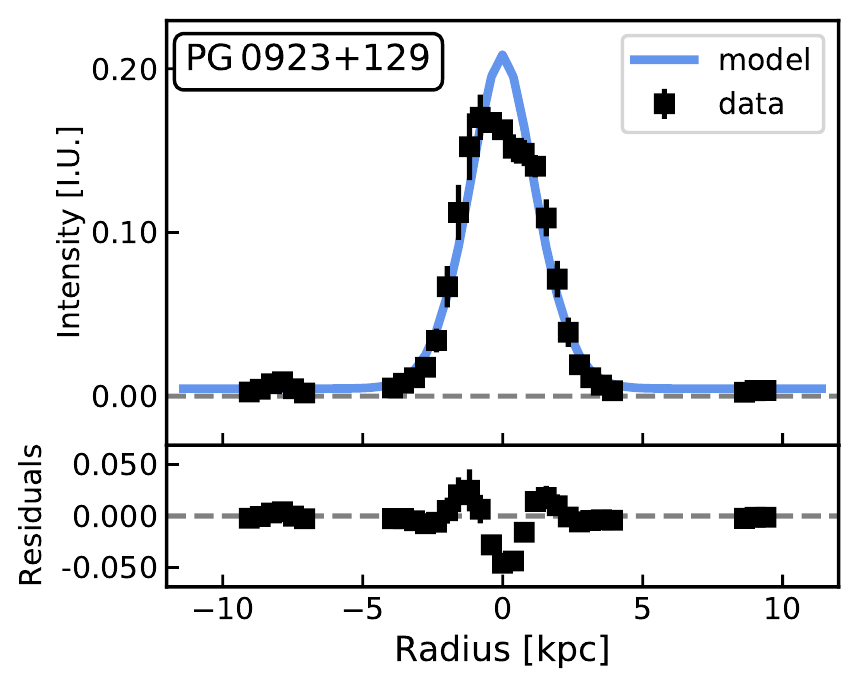}
\includegraphics[width=0.67\columnwidth]{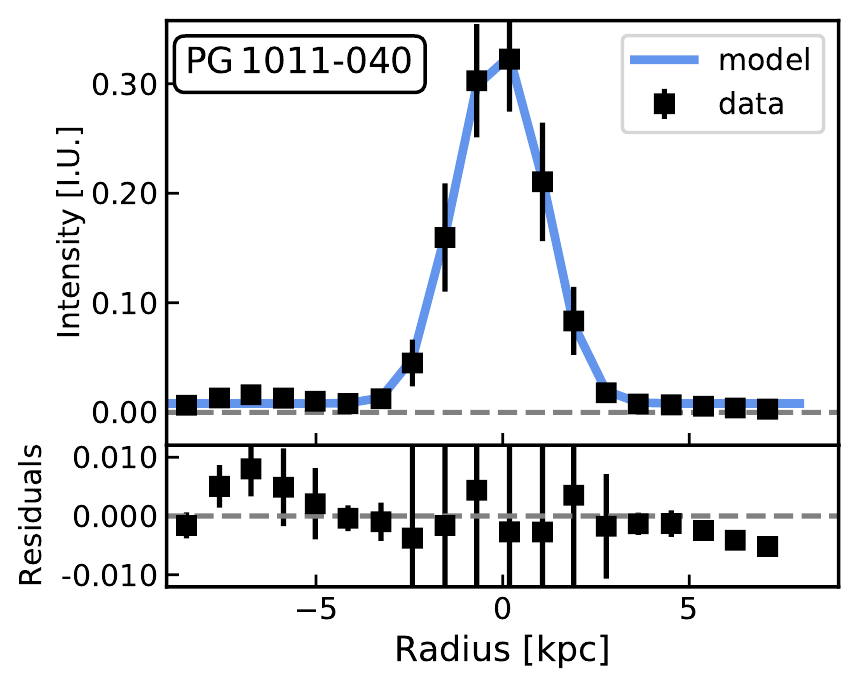}\\
\includegraphics[width=0.67\columnwidth]{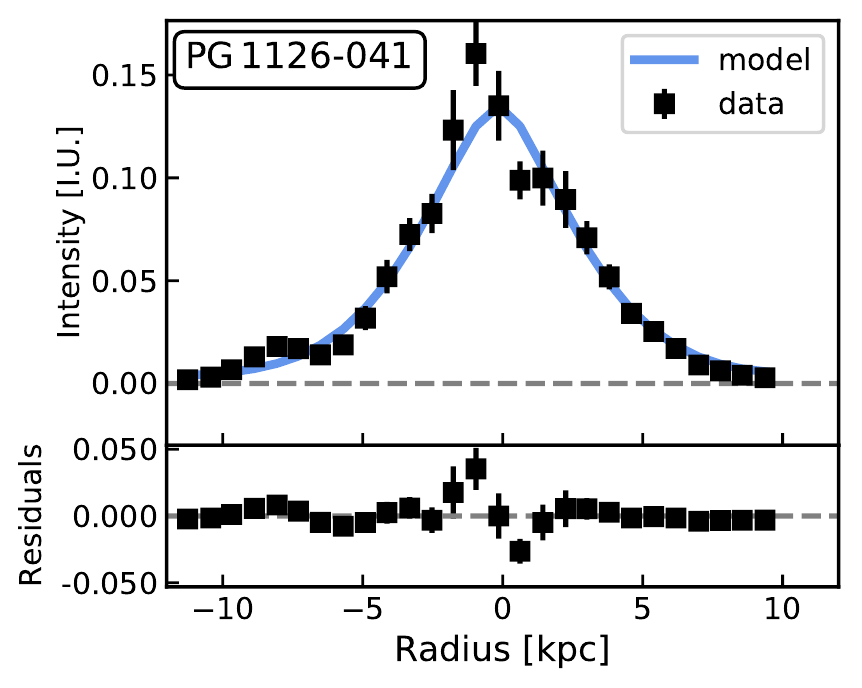}
\includegraphics[width=0.67\columnwidth]{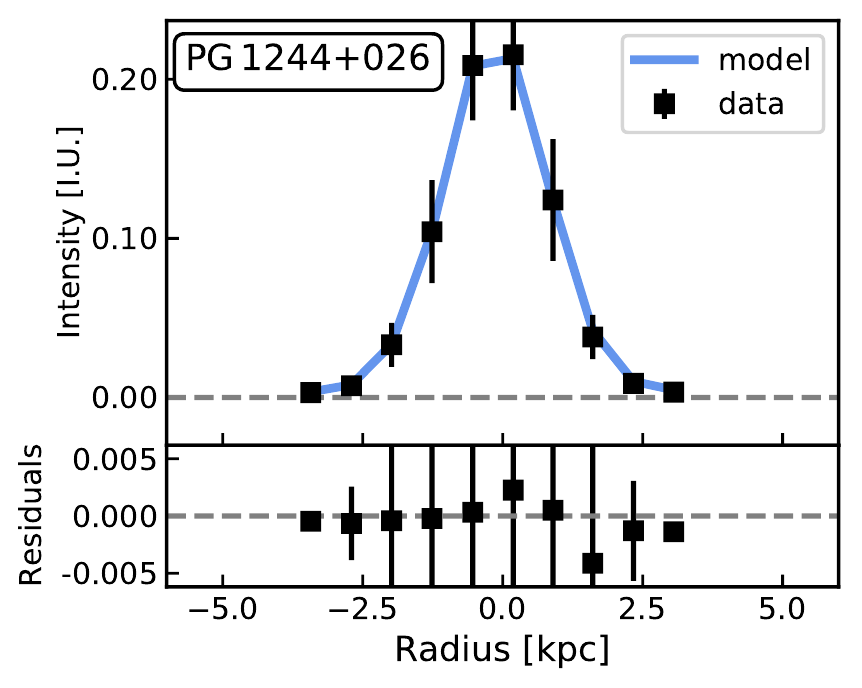}
\includegraphics[width=0.67\columnwidth]{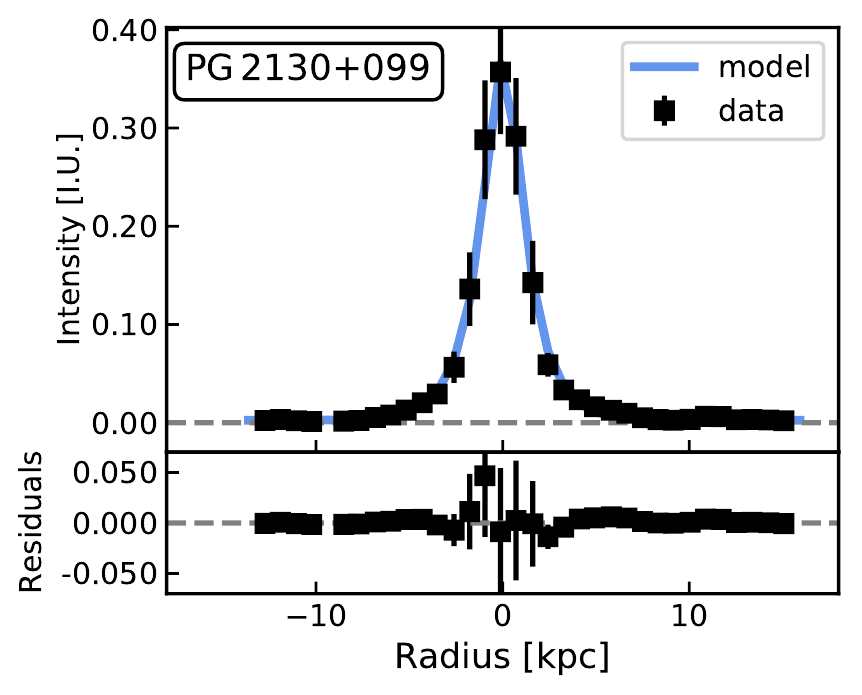}\\
\caption{\label{fig:int_profiles} Intensity profiles across the major axis of the CO(2--1) image, compared with model profiles extracted from the best-fit two-dimensional  S\'ersic models.  We find good agreement between the data and models, as the residual panels suggest.}
\end{figure*}

\subsection{CO(2--1) Intensity Map Modeling}
\label{sec:glob_mod}

We model the two-dimensional intensity maps using a radial profile described by a \cite{Sersic1963} function,

\begin{equation}
I(R) = I_e\,{\rm exp} \left\lbrace - b_{n} \left[ \left( \frac{R}{R_e} \right)^{1/n} - 1 \right] \right\rbrace,
\label{eq:sersic}
\end{equation}

\noindent where $b_{n}$ is a constant that sets $R_e$ as the effective (half-light) radius, $I_e$ is the intensity measured at $R_e$, and $n$ is the S\'ersic index.  An exponential function, which often describes a cold disk, corresponds to $n = 1$.  The two-dimensional model considers seven free parameters ($I_e$, $n$, $R_e$, PA$_{\rm pho}$, $b/a$ minor-to-major axis ratio, $[x_0, y_0]$ on-sky center location, and background) and the S\'ersic model is convolved with the synthesized beam to recover unbiased estimates.

To model the galaxy projection on the sky, we assume that the host galaxies follow an oblate spheroidal geometry \citep{Hubble1926}, such that

\begin{equation}
\cos^2 i =\frac{(b/a)^2-q_0^2}{1-q_0^2},
\label{eq:inc_eqn}
\end{equation}

\noindent where $q_0$ is the intrinsic thickness.  We set $q_0 = 0.14$ based on the mean value reported for edge-on galaxies at nearly the same redshift range ($z<0.05$; \citealt{Mosenkov2015}; see \citealt{Yu2020} for a more complicated prescription).

We use the \textsc{Python} package \textsc{emcee} \citep{Foreman2013} to find the best-fit model. Briefly, \textsc{emcee} implements the affine invariant ensemble sampler for the Markov chain Monte Carlo sampling method to characterize the model probability density function. This allows the algorithm to perform equally well under affine transformations (including linear transformations) between the model parameters, and therefore is less sensitive to the possible covariances among them (see \citealt{Foreman2013} for more details).  We optimize the log-likelihood

\begin{equation}
\ln \mathcal{L} \equiv - \frac{1}{2} \sum_i^N \left( \frac{(y_i - y^m_i)^2}{\sigma_i^2} + \ln(2 \pi \sigma_i^2) \right),
\label{eq:likelihood}
\end{equation}

\noindent where $y_i$ denotes the pixel data taken from the intensity maps, $\sigma_i$ is the $1\,\sigma$ uncertainty, and $y_i^m$ corresponds to the model value for that pixel.

\begin{table*}
	\centering
	\def\arraystretch{1.5}
	\setlength\tabcolsep{3pt}
    	\caption{\label{tab:bestpar} CO(2--1) Emission and Spatial Distribution}
    	\vspace{1mm}
	\begin{tabular}{ccccccccc} 
		\hline
		\hline
		Object & $S_{\rm CO(2-1)} \Delta v$ & $\log L^\prime_{\rm CO(2-1)}$ & $I_e$ & $n$ & $R_e$ & $b/a$ & PA$_{\rm pho}$ & $R_{\rm 1/2}$ \\
 		& (Jy\,km\,s$^{-1}$) & (K\,km\,s$^{-1}$\,pc$^2$) & (Jy\,km\,s$^{-1}$) & & (kpc) & & ($^\circ$) & (kpc)\\
 		(1) & (2) & (3) & (4) & (5) & (6) & (7) & (8) & (9)\\
		 \hline
		PG\,0050+124    & $71.32\pm0.08$ & $9.515\pm0.001$ & $0.020 \pm 0.001$ & $1.89 \pm 0.02$ & $1.13 \pm 0.01$ & $0.86 \pm 0.01$ & $15 \pm 1$ & $1.4\pm 0.1$\\
		PG\,0923+129    & $42.38\pm0.06$ & $8.637\pm0.001$ & $0.033 \pm 0.001$ & $0.58 \pm 0.01$ & $1.52 \pm 0.01$ & $0.74 \pm 0.01$ & $53 \pm 1$ & $1.8 \pm 0.2$\\
		PG\,1011$-$040  & $18.90\pm0.04$ & $8.894\pm0.001$ & $0.099 \pm 0.002$ & $0.44 \pm 0.01$ & $1.29 \pm 0.01$ & $0.81 \pm 0.01$ & $102 \pm 1$ & $1.8 \pm 0.4$\\
		PG\,1126$-$041  & $17.56\pm0.04$ & $8.892\pm0.001$ & $0.018 \pm 0.001$ & $0.70 \pm 0.03$ & $3.79 \pm 0.01$ & $0.35 \pm 0.01$ & $322 \pm 1$ & $3.7 \pm 0.3$\\
		PG\,1244+026    & $6.06\pm0.02$   & $8.234\pm0.001$ & $0.057 \pm 0.001$ & $0.53 \pm 0.01$ & $1.07 \pm 0.01$ & $0.83 \pm 0.01$ & $186 \pm 1$ & $0.9 \pm 0.3$\\
		PG\,2130+099    & $16.35\pm0.04$ & $8.876\pm0.001$ & $0.044 \pm 0.001$ & $1.07 \pm 0.01$ & $1.89 \pm 0.01$ & $0.67 \pm 0.01$ & $44 \pm 1$  & $1.8 \pm 0.4$\\		
		\hline                                                                           
	\end{tabular}
	\justify	
	 {\textsc{Note}--- (1) Source name.  (2) Velocity-integrated flux density estimated from our concatenated data. We have not considered the ALMA flux calibration uncertainty ($\lesssim 10$\,\%; \citealt{Fomalont2014,Bonato2018}). (3) CO(2--1) line luminosity, updated from the concatenated data measurements. (4) Velocity-integrated line intensity at the effective radius. (5) S\'ersic index. (6) Effective radius. (7) Ratio of the projected minor axis to to major axis. (8) Position angle (North\,$= 0^\circ$, East\,$= 90^\circ$). (9) CO(2--1) half-light radius calculated by implementing a tilted-ring approach and using the best-fit PA$_{\rm pho}$ and $b/a$ values (Col. 7 and 8).}
\end{table*}

The intensity maps, along with the residuals from the best-fit model, are shown in Figure~\ref{fig:int_maps}, while the best-fit parameters are presented in Table~\ref{tab:bestpar}.  These best-fit parameters are also used to extract the radial intensity profile for each system (Figure~\ref{fig:int_profiles}), which is derived by simulating a slit with width equal to the beam FWHM and aligned with respect to the major axis of the CO(2--1) image (given by PA$_{\rm pho}$). Across the simulated slit the data are sampled in radial bins with size equal to half of the beam FWHM to avoid over- and under-sampling. For each radial bin, we calculate the average intensity value and the $1\,\sigma$ uncertainty, which is given by the standard deviation of the encompassed individual pixel values.

The stellar component of the six quasar host galaxies are all observed by HST with high spatial resolution at optical or NIR wavelengths.  HST images of PG\,0050+124, PG\,0923+129, PG\,1011$-$040, and PG\,1244+026 are taken from WFC3 camera in rest-frame $I$-band (PI: L. C. Ho).  Images of PG\,1126$-$041 and PG\,2130+099 are taken by NICMOS ($H$-band; PI: S. Veilleux) and WFPC2 ($B$-band; PI: S. R. Heap).  We collect them in Figure~\ref{fig:int_maps} as a reference for the CO emission. The molecular gas tends to be distributed within the stellar component delineated by the HST images.  

In four sources (PG\,0050+124, PG\,0923+129, PG\,1126$-$041, PG\,2130+099), CO emission is detected as far out as $\sim 10$\,kpc from the center, although the bulk of it is confined to a more compact, disk-like morphology.  The CO morphology of PG\,1244+026 is the most compact, with a total radial extension $\lesssim 4$\,kpc (Figure~\ref{fig:int_profiles}). PG\,1011$-$040 exhibits a complex structure. Appendix~\ref{sec:AppA} gives comments on the morpho-kinematics of each source.

Despite the observed variety of CO morphologies, the two-dimensional flux distributions are well fitted by S\'ersic models with indexes in the range of $n \approx 0.4-1.9$.  Subtracting the best-fitting global component from each map reveals complex sub-structures in the residual maps, ranging from clumps (PG\,1126$-$041 and PG\,1011$-$040) to inner spiral arms (PG\,0923+129 and PG\,1126$-$041).  PG\,0923+129 and PG\,2130+099 further show a central cavity surrounded by a ring-like structure. However, we caution that PG\,0923+129 seems to present a central plateau in its CO surface brightness distribution, as can be seen from its radial profile (Figure~\ref{fig:int_profiles}). It is possible that the apparent central depression is produced by model over-subtraction.

\begin{figure}
\centering
\includegraphics[width=1.0\columnwidth]{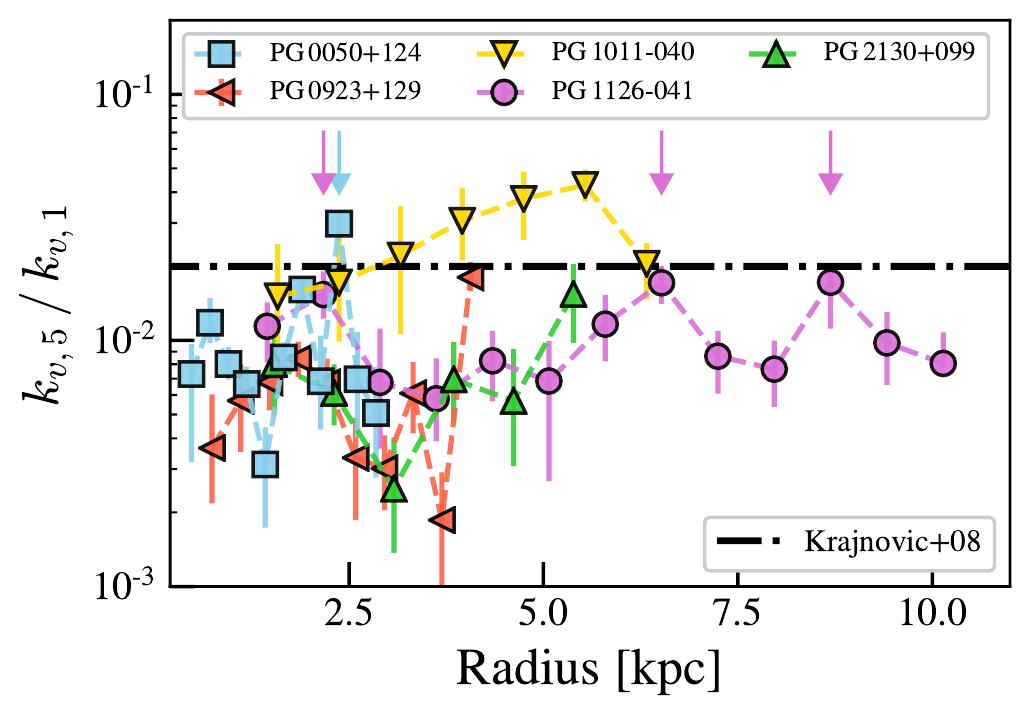}\\
\caption{\label{fig:kinemetry} Radial variation of the \textsc{kinemetry} coefficient $k_{v,5}/k_{v,1}$ for five PG quasar host galaxies. The dot-dashed line corresponds to $\langle k_{v,5}/k_{v,1} \rangle = 0.02$ threshold that separates between ``regular'' and ``non-regular'' rotators \citep{Krajnovic2008}. The colored arrows indicate the host galaxy radius at which $k_{v,5}/k_{v,1}$ is consistent with the threshold value. We find that four host galaxies are classified as regular rotators, while only one system, PG\,1011$-$040, displays perturbed molecular gas kinematics.}
\end{figure}

\subsection{Kinemetry}
\label{sec:kinemetry}

We investigate whether the observed molecular gas rotational motions resemble those expected from an ideal disk by analyzing the two-dimensional LOS velocity maps using the tilted-ring approach as implemented in \textsc{kinemetry} \citep{Krajnovic2006}\footnote{http://davor.krajnovic.org/idl/}.  \textsc{kinemetry} quantifies the deviations from disk-like kinematics for an observed velocity field by parameterizing it as a function of the radius ($R$) and the azimuthal angle ($\psi$). For an ideal disk, the velocity profile is only a function of the galaxy radius, and the sky projection adds a variation in terms of azimuthal angle that follows a cosine law: $v(R, \psi) = v(R) \cos (\psi)$.  To test this condition, \textsc{kinemetry} decomposes the LOS velocity into a series of tilted rings. Along each elliptical path, it parameterizes the velocity profile in terms of a Fourier series that only depends on the azimuthal angle: 

\begin{equation}
\label{eq:kinemetry}
v_{\rm LOS}(a,\psi) = A_0(a) +  \sum^N_{n=1} k_{v,n}(a) \cos[n(\psi-\phi_n(a))],
\end{equation}

\noindent
where $k_{v,n}$ and $\phi_n$ are the amplitude and phase coefficients, and $a$ is the length of the kinematic major axis. The kinematic position angle (PA$_{\rm kin}$) is accurately recovered in tilted rings dominated by a single component, whereas in the ellipses between multiple components it traces the position of maximum velocity amplitude. The case of an ideal disk is recovered when all the $k_{v,n}$ coefficients are zero except $k_{v,1}$. Hence, the high-order amplitude coefficients quantify the kinematic deviations or ``asymmetries'' observed in the velocity maps (see \citealt{Krajnovic2006} for more details).

\textsc{kinemetry} does not fit the kinematic center, and this parameter must be given in advance. We use the center determined by the best-fit S\'ersic model as input. We also fix the axis ratio equal to the value given by our morphological modeling, as this parameter is not well-constrained by fitting the LOS velocity map alone.  The tilted ring thickness is set equal to half of the beam FWHM, and we skip the first tilted ring iteration in each map because the radius is smaller than the beam FWHM.  We restrict the analysis to the central zone of the galaxy, which, in any case, is our primary interest.  We find that \textsc{kinemetry} does not interpolate reliably the maps in the outer regions, where spiral arms may be present (e.g., PG\,0050+124), producing spurious results such as inverted kinematic position angles at large radii.  This restriction also helps to avoiding outer zones that may be under-sampled due to incomplete azimuthal coverage. For each tilted ring, we use the default 75\% pixel sampling limit required by \textsc{kinemetry} to obtain reliable estimates \citep{Krajnovic2006}. 

To consider the uncertainties associated with the kinematic center and the LOS velocity map, we bootstrap these values within their $3\,\sigma$ error range and repeat the \textsc{kinemetry} procedure 100 times. Then, the $1\,\sigma$ uncertainties of the $k_{n,v}$ amplitude coefficient are estimated from the corresponding distribution by calculating the 16th and 84th percentiles.

We quantify the non-regular motions by calculating the ratio $k_{5,v}/k_{1,v}$ (e.g., \citealt{Krajnovic2008,Krajnovic2011,vandenSande2017}). The fifth-order amplitude coefficient is used because \textsc{kinemetry} adjusts the lower order coefficients to find the best-fit tilted ring. \citet{Krajnovic2011} used an average threshold ratio of $\langle k_{5,v}/k_{1,v} \rangle = 0.04$ to classify galaxies as ``regular'' or ``non-regular'' rotators for the the ATLAS$^{\rm 3D}$ survey \citep{Cappellari2011}. However, this threshold depends on data quality measured by the typical uncertainty in $k_{5,v}/k_{1,v}$. For example, \cite{Krajnovic2008} used a limit of $\langle k_{5,v}/k_{1,v} \rangle = 0.02$ to determine if a galaxy presents `disk-like rotation' for the SAURON project \citep{deZeeuw2002}. Based on the typical uncertainty of $k_{5,v}/k_{1,v}$ measured for our systems ($\sim 0.001$), we adopt a threshold of $\langle k_{5,v}/k_{1,v}\rangle = 0.02$ for this study.  We apply the \textsc{kinemetry} analysis for five of the six PG quasars\footnote{We avoid analyzing PG\,1244$+$026 because of the high degree of compactness of the system. With a projected major axis extension of $\sim 7\arcsec$, this source is only sampled by $\sim $5 independent regions across this direction ($\sim 3$ independent tilted rings for characterizing the velocity map), given its beam size of $\sim 1\farcs4$ (Table~\ref{tab:obs}). The effect of beam-smearing may artificially erase any intrinsic kinematic perturbation in this observation, rendering the observed velocity map more ``disky'' than it actually is (e.g., \citealt{Bellocchi2012}).}.  The radial median of $k_{v,5}/k_{v,1}$ spans $0.006-0.027$ (Table~\ref{tab:kinpar}), such that four would be classified as regular rotators, with $k_{v,5}/k_{v,1} < 0.02$ over almost all radii, with no clear trend toward the inner or outer regions.  PG\,0050+124 and PG\,1126$-$041 have local values of $k_{v,5}/k_{v,1} \approx 0.02$ (as indicated by the arrows in Figure~\ref{fig:kinemetry}), but they are localized galaxy sub-structures observed in the intensity residual maps.  Only PG\,1011$-$040 shows $k_{v,5}/k_{v,1} > 0.02$ at $R \gtrsim 2.5$\,kpc, qualifying it as perturbed.

\begin{figure}
\centering
\includegraphics[width=1.0\columnwidth]{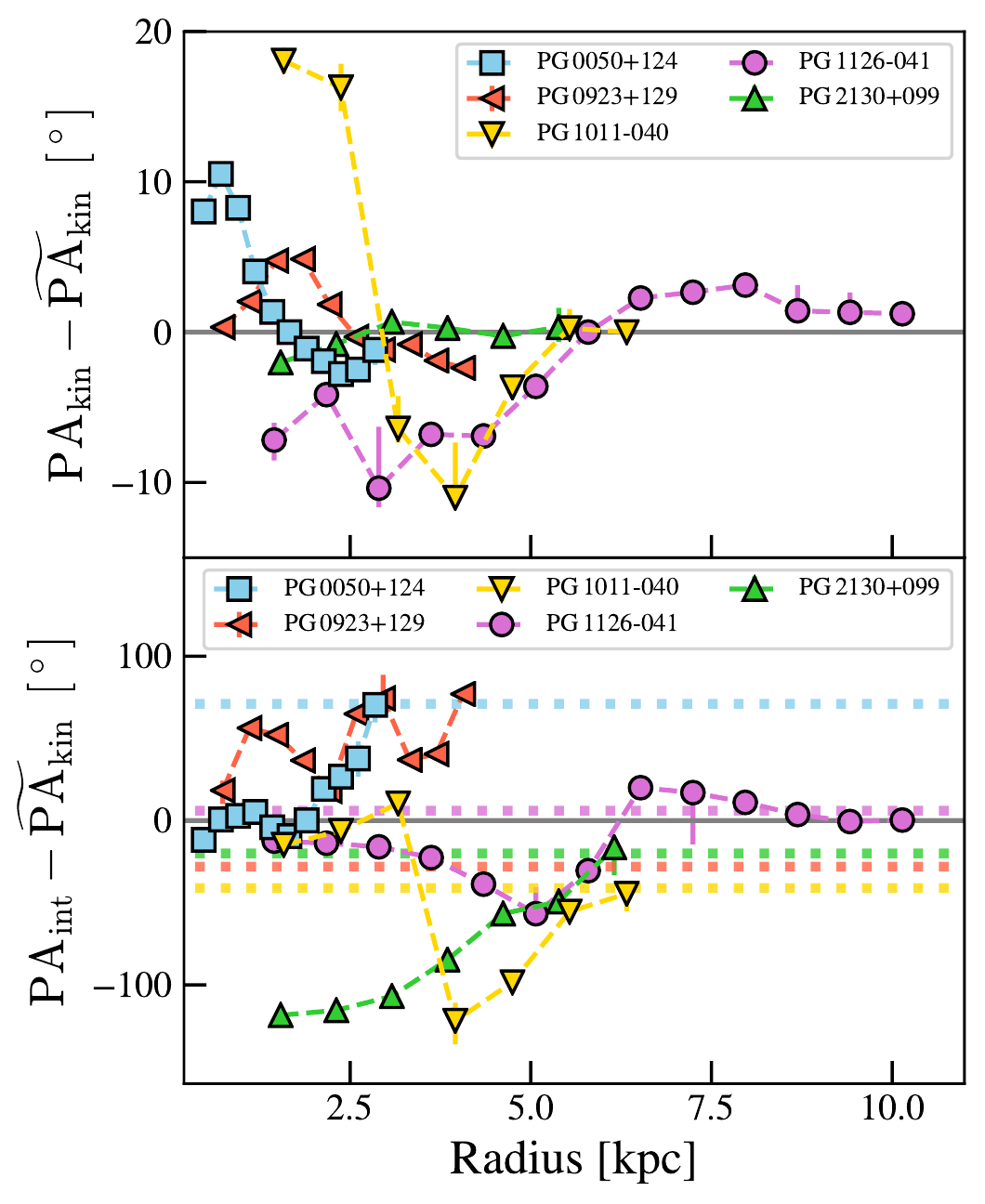}\\
\caption{\label{fig:kinemetry_pa} Radial variation of the position angle measured from the LOS velocity maps (${\rm PA}_{\rm kin}$; \textit{Top}) and intensity maps (${\rm PA}_{\rm int}$; \textit{Bottom}) for five PG quasars. In both panels, the PA values are presented with respect to the median kinematic position angle $\widetilde{\rm PA}_{\rm kin}$. The data are color-coded following Figure~\ref{fig:kinemetry}. The horizontal dotted lines indicate the difference of the global PA$_{\rm pho}$ value with respect to $\widetilde{\rm PA}_{\rm kin}$.}
\end{figure}

The top panel of Figure~\ref{fig:kinemetry_pa} shows the variation of the kinematic position angle (${\rm PA}_{\rm kin}$) with respect to the median value ($\widetilde{\rm PA}_{\rm kin}$) and as a function of radius for the host galaxies. We find that ${\rm PA}_{\rm kin}$ varies smoothly with radius in each source, with a maximum difference of $\Delta {\rm PA}_{\rm kin} \approx 30^\circ$ for PG\,1011$-$040. Three of the five systems (PG\,0050+124, PG\,1011$-$040, PG\,1126$-$041) present a ``kinematic twist'', that is, $\Delta {\rm PA}_{\rm kin} > 10^\circ$ according to the criteria of \cite{Krajnovic2008}.  For each galaxy, we assume that the direction of the kinematic major axis is given by $\widetilde{\rm PA}_{\rm kin}$ (Table~\ref{tab:kinpar}).  The $\widetilde{\rm PA}_{\rm kin}$ values differ on average by $\sim 30^\circ$ from the PA$_{\rm pho}$ values determined by the S\'ersic models. PG\,0050+124 shows the higher difference between both estimates ($\widetilde{\rm PA}_{\rm kin} - {\rm PA}_{\rm pho} \sim 71^\circ$), whereas PG\,1126$-$041 presents the smaller difference ($\widetilde{\rm PA}_{\rm kin} - {\rm PA}_{\rm pho} \sim 4^\circ$). Similar differences are found when we compare $\widetilde{\rm PA}_{\rm kin}$ with the stellar photometric position angles measured from the HST images \citep{Veilleux2009,Yulin2020}. In this case, we find differences in the range of $0 - 132^\circ$, with an average difference of $\sim 25^\circ$.

To understand the difference between $\widetilde{\rm PA}_{\rm kin}$ and PA$_{\rm pho}$ values, we also analyze the PG quasars CO(2--1) intensity maps using \textsc{kinemetry}.  In this case, \textsc{kinemetry} performs a simple ellipse fitting method.  We apply the same restrictions that we assumed for the LOS velocity map analyzes. In the bottom panel of Figure~\ref{fig:kinemetry_pa} we show the difference of the position angles derived from the intensity maps (PA$_{\rm int}$) with respect to $\widetilde{\rm PA}_{\rm kin}$. We measure larger radial variations of PA$_{\rm int}$ ($\Delta {\rm PA}_{\rm int} \approx 130^\circ$) when compared to ${\rm PA}_{\rm kin}$ ($\Delta {\rm PA}_{\rm kin} \approx 30^\circ$), highlighting the effect of molecular gas sub-structure on the intensity maps, but only relatively subtle fingerprints on the LOS velocity fields. With the exception of PG\,0923+129, PA$_{\rm int}$ approximates to PA$_{\rm pho}$ (horizontal dotted lines in Figure~\ref{fig:kinemetry_pa}) at longer radii. On the other hand, PA$_{\rm int}$ tends to be consistent with $\widetilde{\rm PA}_{\rm kin}$ at small radii (except for PG\,2130+099). The difference between $\widetilde{\rm PA}_{\rm kin}$ and PA$_{\rm pho}$ may be produced by the large variation of the photometric position angle values tracing multiple molecular gas components induced by, perhaps, stellar bars (e.g., PG\,1011$-$040).

\begin{figure}
\centering
\includegraphics[width=0.72\columnwidth]{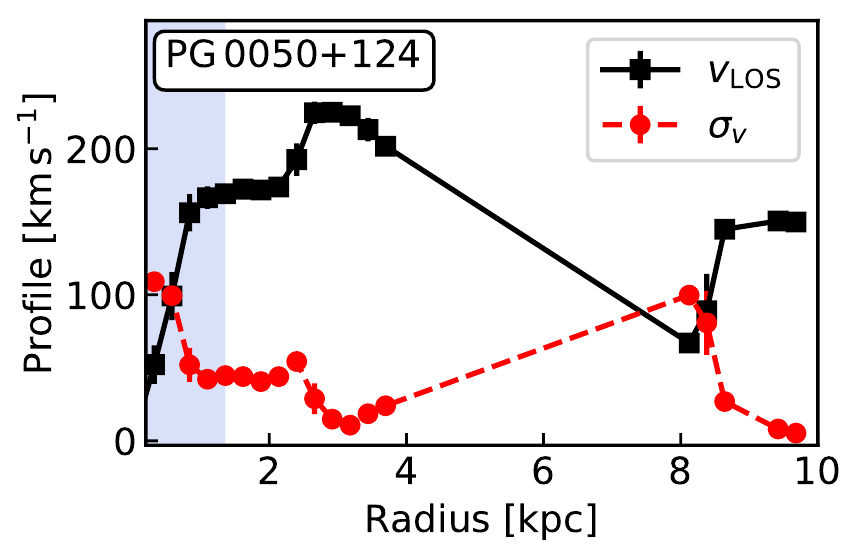}\\
\includegraphics[width=0.72\columnwidth]{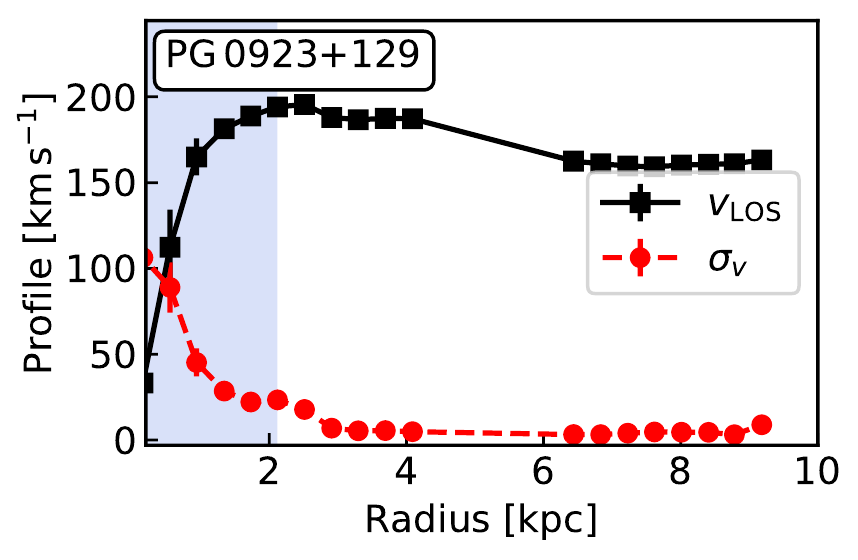}\\
\includegraphics[width=0.72\columnwidth]{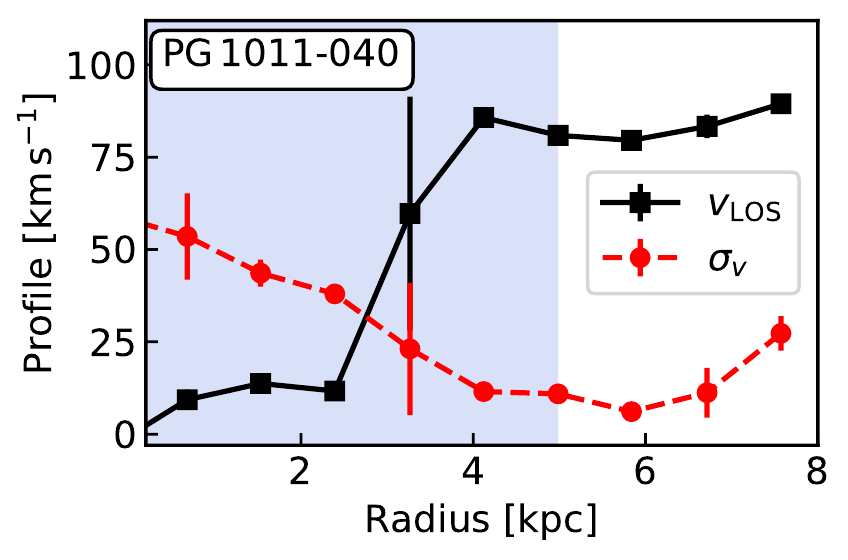}\\
\includegraphics[width=0.72\columnwidth]{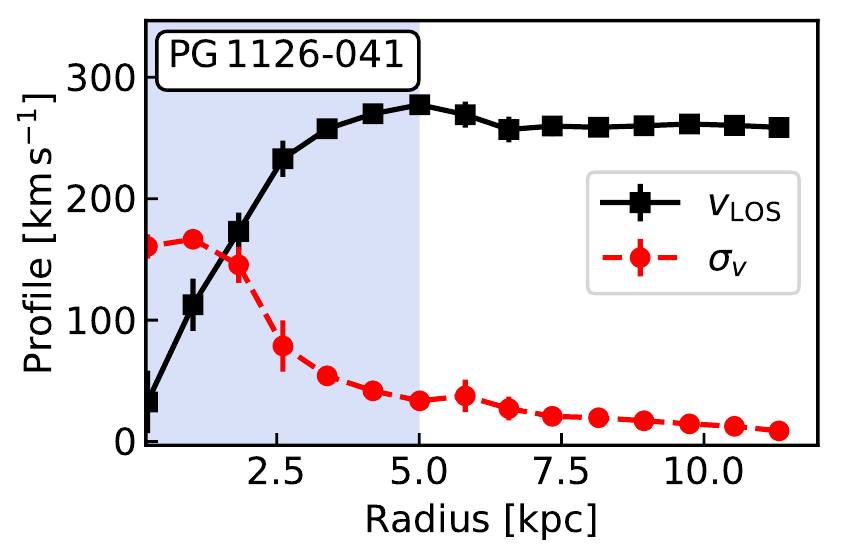}\\
\includegraphics[width=0.72\columnwidth]{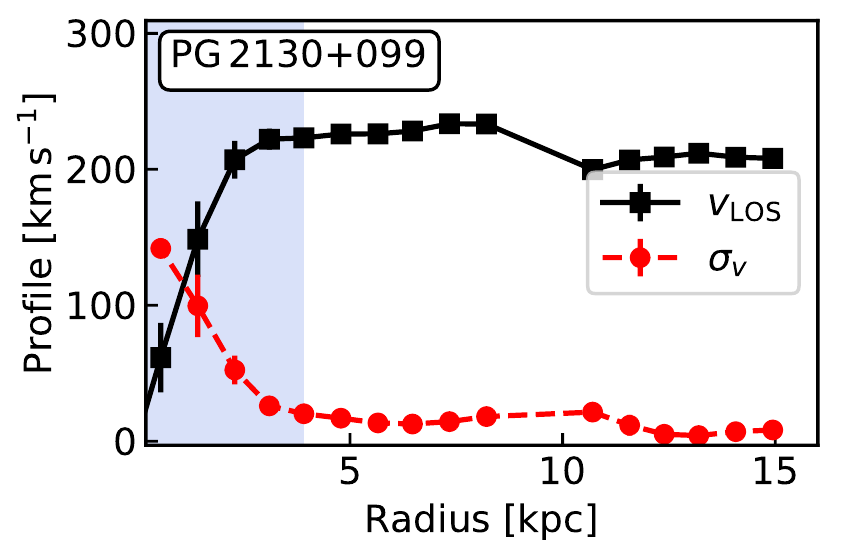}\\
\caption{\label{fig:vel_profiles} LOS velocity and velocity dispersion profiles across the kinematic major axis for five PG quasars. The shaded zones represent the inner regions that suffer from significant beam-smearing effect.  PG\,0050$+$124 shows considerably elevated $\sigma_v$ values in its central zone ($R \lesssim 2.5$\,kpc; see also Figure~\ref{fig:int_maps}). PG\,1244$+$026 has no profile because we do not have an estimate of its kinematic major axis.}
\end{figure}

We also try to identify gas elliptical streaming or radial flow signatures across the host galactic disks (e.g. \citealt{Wong2004}).  Briefly, by writing Equation~\ref{eq:kinemetry} as an harmonic sum of sines and cosines (e.g. \citealt{Schoenmakers1997}), the 1st and 3rd order coefficients multiplying the sine terms ($s_1$ and $s_3$, respectively) may be indicative of elliptical streaming or radial flows.  Elliptical streaming generally produces an anti-correlation between the $s_1$ and $s_3$ terms, while axisymmetric radial flows can be identified in the case of significant $s_1$ value but negligible $s_3$ term ($| s_1 / s_3 | \gtrsim 10$). However, a static bar potential can produce the same kinematic signature of a radial flow (see \citealt{Wong2004}, for more details).  We obtain the $s_1$ and $s_3$ radial profiles directly from \textsc{kinemetry} for the host galaxies. The $s_1$ and $s_3$ coefficients absolute values are within $\lesssim 7$\,km\,s$^{-1}$, with no clear evidence of gas elliptical streaming or axisymmetric radial flows in any host galaxy.  It is worth to mention that results obtained from the harmonic decomposition analyses of the velocity fields must not be considered as conclusive. Kinematic features produced by non-axysimmetric sub-structures in disk galaxies limitate further interpretation \citep{Wong2004}. Moreover, radial flows signatures suggested by gravitational torque modeling do not necessarily coincide with the expectation from the purely kinematic decomposition analysis \citep{Haan2009}. Any interpretation of the observed kinematics requires a detailed knowledge of the galaxy potential (e.g. \citealt{Alonso-Herrero2018}).

\subsection{Regular Motions and Velocity Dispersion}
\label{sec:reg_res}

Rotational patterns can be seen in the LOS velocity maps, while velocity dispersion rises toward the center (Figure~\ref{fig:int_maps}).  To further study the kinematics, we extract velocity radial profiles along the kinematic major axis to avoid any effect produced by the disk azimuthal projection. These velocity radial profiles, available for five of the six sources (Figure~\ref{fig:vel_profiles}), are derived in the same manner as the intensity radial profiles, but we simulate a slit aligned with the kinematic major axis instead of the photometric major axis.  We can measure rotation curves to $\gtrsim 8$\,kpc in all cases, and the flat part of the rotation curve is reached for at least PG\,0923+129, PG\,1126$-$041, and PG\,2130+099. 

We calculate the local projected velocity gradients from the rotation curves.  Prior to estimating representative rotation velocities for each system, it is first necessary to avoid the central region of the galaxy, where the LOS velocities are biased toward lower values due to beam-smearing. From an observational point of view, beam-smearing effects are minimized in regions where the local projected velocity gradient is smaller than the spectral resolution.  Under these conditions, the line centroids are located at nearly the same spectral channel, and the rotation velocities can be recovered directly from the data.  Thus, we mask the data from the nucleus to the radius where the local velocity gradient decreases below the spectral resolution limit of $11$\,km\,s$^{-1}$; the masked region is denoted by the shaded blue area in Figure~\ref{fig:vel_profiles}.  Representative values of the rotation velocity ($v_{\rm rot}$; Table~\ref{tab:kinpar}) are estimated by the mean value of the unmasked data points on the rotation curve, and the $1\,\sigma$ uncertainty is estimated from their standard deviation. We correct these values for inclination projection. An analogous treatment is applied to the velocity dispersion ($\sigma_v$), using the same mask applied to the rotation velocity curve so as to avoid the central pixels affected by beam-smearing (e.g., \citealt{Davies2011,Wisnioski2015,Stott2016}). 

We find $v_{\rm rot} / \sigma_v = 9$--42 for the five analyzed host galaxies, indicating that the gas kinematics are dominated by rotation. This is consistent with the range measured for the EDGE-CALIFA galaxies ($v_{\rm rot} / \sigma_v = 10$--28; \citealt{Levy2018}).  However, different from the EDGE-CALIFA systems, $\sigma_v$ varies considerably from object to object. For example, the average dispersion of PG\,0050+124 ($\sigma_v = 36\pm27$\,km\,s$^{-1}$) is 6 times higher than that of PG\,0923+129 ($\sigma_v = 6\pm4$\,km\,s$^{-1}$).  The average $\sigma_v$ values reported for the EDGE-CALIFA galaxies are in the range of $\sim 9-19$\,km\,s$^{-1}$ \citep{Levy2018}.  The average value of PG\,0923+129 is consistent with the estimate traditionally adopted for nearby star-forming galaxies ($\sim 6$\,km\,s$^{-1}$; \citealt{Leroy2008}).  On the other hand, PG\,0050+124 presents an average $\sigma_v$ value larger than any of those reported for the EDGE-CALIFA systems. These values are calculated over portions of the dispersion velocity profile that should be immune from beam-smearing, and thus these may reflect different physical properties of each host galaxy or degree of AGN activity. Detailed study of the molecular gas velocity dispersion will be addressed in a future work.
 
\begin{table}
	\centering
	\def\arraystretch{1.0}
	\setlength\tabcolsep{3pt}
    	\caption{\label{tab:kinpar} Kinematic Parameters}
    	\vspace{1mm}
	\begin{tabular}{ccccc} 
		\hline
		\hline
		Object & $\widetilde{\rm PA}_{\rm kin}$ & $k_{v,5} / k_{v,1}$ & $v_{\rm rot}$ & $\sigma_v$  \\
 		& ($^\circ$) & & (km\,s$^{-1}$) & (km\,s$^{-1}$) \\
 		(1) & (2) & (3) & (4) & (5) \\
		 \hline
		PG\,0050+124    & $304$ & 0.010 & $337 \pm 92$ & $36 \pm 27$\\
		PG\,0923+129    & $81$ & 0.006 & $253 \pm 20$ & $6 \pm 4$\\
		PG\,1011$-$040 & $151$ & 0.027 & $141 \pm 7$ & $15 \pm 9$\\
		PG\,1126$-$041 & $326$ & 0.010 & $275 \pm 4$ & $20 \pm 9$\\
		PG\,2130+099    & $64$ & 0.007 & $290 \pm 15$ & $12 \pm 5$\\		
		\hline                                                                           
	\end{tabular}
	\justify	
	 {\textsc{Note}--- (1) Source name. (2) Median position angle of the kinematic major axis (with respect to the receding side) measured from \textsc{kinemetry} analysis (North\,$= 0^\circ$, East\,$= 90^\circ$). (3) Median value of the ratio of the fifth-order amplitude coefficient over the first-order coefficient; the typical $1\,\sigma$ uncertainty is 0.001. (4) Average rotation velocity derived from the rotation curve, corrected for inclination. (5) Average velocity dispersion derived from the velocity dispersion profile. \par}
\end{table}

\begin{figure*}
\centering
\includegraphics[width=1.5\columnwidth]{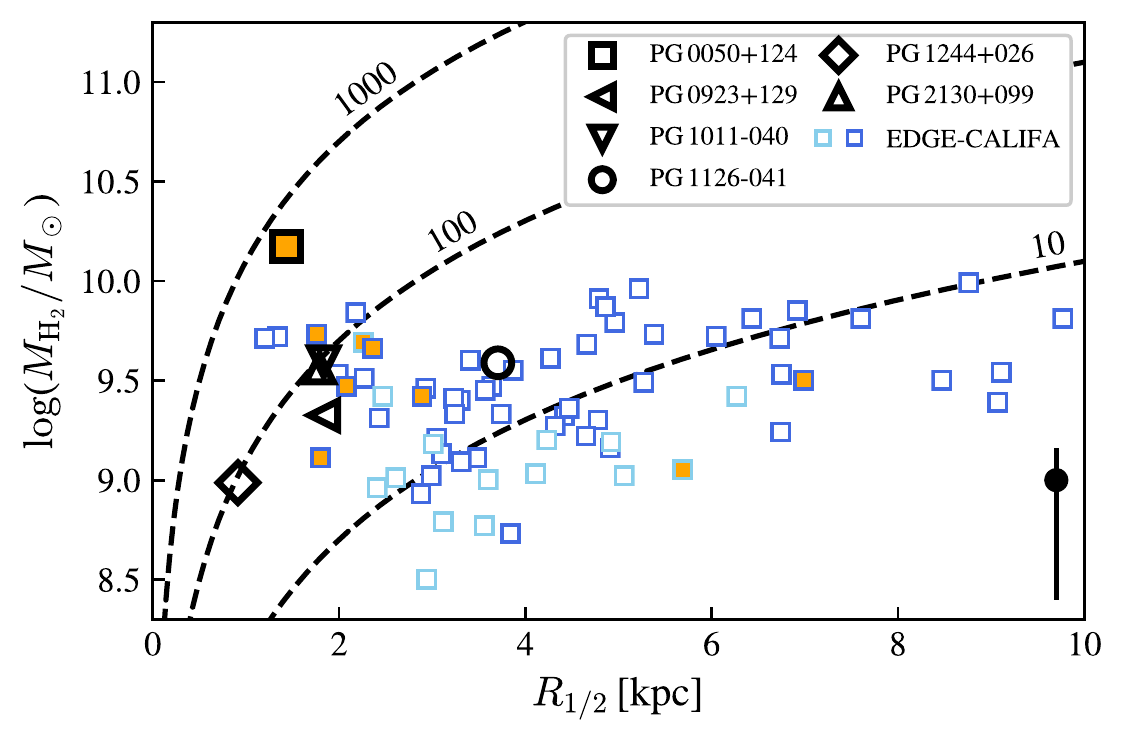}\\
\caption{\label{fig:COsize_comp} Distribution of molecular gas mass versus the half-light radius of the CO distribution, as measured for the six PG quasars and the star-forming galaxies in the EDGE-CALIFA survey. The EDGE-CALIFA data are presented in light blue and blue squares, where the latter symbol represents the subsample that better matches the PG quasars in terms of SFR, $M_\star$, $M_{\rm H_2}$, and $f_{\rm H_2}$. The dashed curves correspond to values of constant molecular gas surface density measured in [$M_\odot$\,pc$^{-2}$] units. The black bar in the lower-right corner illustrates the variation in $M_{\rm H_2}$ induced by the choice of $\alpha_{\rm CO}$: the range above and below the black dot represent the effect of varying  $\alpha_{\rm CO}$ between the value for the Milky Way and ultra-luminous infrared galaxies. The filled orange symbols indicate the galaxies that belong to a multiple system. The observed PG quasars tend to have more compact molecular gas distributions compared to the EDGE-CALIFA galaxy subsample.}
\end{figure*}

\subsection{Compact Molecular Gas Distribution}
\label{sec:comparison_res}

In terms of global quantities, the interstellar medium content of the PG quasars seems to be similar to that of normal star-forming galaxies of the same stellar mass \citep{Shangguan2018,Shangguan2020}. This is emphasized again in Figure~\ref{fig:comparison_sample}, where we show that the majority of the host galaxies with CO measurements have similar $M_{\rm H_2}$, $f_{\rm H_2}$, $M_\star$, and SFR relative to inactive galaxies observed by the xCOLD\,GASS \citep{Saintonge2017} and EDGE-CALIFA surveys \citep{Bolatto2017}. However, one missing key information is the spatial distribution of the molecular gas, which now can be investigated with our new observations.

Figure~\ref{fig:COsize_comp} shows the molecular gas content of the PG quasars plotted as a function of the half-light radius ($R_{1/2}$) measured from the intensity maps by implementing a tilted ring approach using the best-fit S\'ersic model parameters. We use $R_{\rm 1/2}$ instead of $R_e$ for consistency with the methodology employed by \cite{Bolatto2017} for the EDGE-CALIFA galaxies, which serve as a direct comparison.\footnote{The $R_{\rm 1/2}$ estimates agree with the $R_e$ values for the six host galaxies (Appendix~\ref{sec:AppB}).}  The blue squares in Figure~\ref{fig:COsize_comp} highlight the subset of EDGE-CALIFA galaxies that better matches the quasars in terms of their global physical properties (Figure~\ref{fig:comparison_sample}).  

The six quasar hosts tend to have compact molecular gas distributions: $R_{\rm 1/2} = 0.9 - 3.7$\,kpc, with a median value of 1.8\,kpc.  Defining the molecular gas surface density as $\Sigma_{\rm H_2}  \equiv M_{\rm H_2} / \pi (2 R_{1/2})^2$, the sample is characterized by $\Sigma_{\rm H_2} \gtrsim 22 $\,$M_\odot$\,pc$^{-2}$ (see dashed curves in Figure~\ref{fig:COsize_comp}). The PG quasars tend to be located at the high-$\Sigma_{\rm H_2}$, small-$R_{1/2}$ region of the CO mass-size plane compared to the EDGE-CALIFA galaxies.  We note, however, that a minority of inactive galaxies also have comparable $\Sigma_{\rm H_2}$ and $R_{1/2}$, and compact molecular gas distribution is not a property unique to AGNs.\footnote{The five AGN hosts previously discarded from the EDGE-CALIFA sample (not shown in Figure.~\ref{fig:COsize_comp}) have $R_{1/2} \sim 3-7$\,kpc and  $\Sigma_{\rm H_2} \sim 10 $\,$M_\odot$\,pc$^{-2}$.}  We further distinguish between the galaxies that belong to a multiple system (filled orange symbols). This is based on the visual detection of any system (independent of its size or brightness) within the HST image field-of-view for the PG sources (Figure~\ref{fig:int_maps}) and literature morphological classifications of the EDGE-CALIFA galaxies, assembled by \cite{Bolatto2017}.  As \cite{Bolatto2017} have noted,  EDGE-CALIFA  galaxies in a multiple system tend to be more compact.  The sample of quasars (6 out of 40 sources at $z <0.3$) is too small to draw any meaningful conclusions in this regard.  

Our statements on $\Sigma_{\rm H_2}$ obviously depend on the choice of $\alpha_{\rm CO}$.  This is especially worrisome if $\alpha_{\rm CO}$ drops in environments of high surface density \citep{Bolatto2013}.  \cite{Shangguan2020}, in comparing CO observations of the PG quasars with independent gas masses inferred from far-infrared dust emission, find no evidence that quasar host galaxies possess abnormal values of $\alpha_{\rm CO}$ as it is found for luminous and ultra luminous infrared galaxies ($\sim 1.8$\,$M_\odot$\,(K\,km\,s$^{-1}$\,pc$^2$)$^{-1}$; \citealt{Herrero-Illana2019}). In any event, our main conclusions do not qualitatively change within the likely range of $\alpha_{\rm CO}$ (0.8--4.36\,$M_\odot$\,(K\,km\,s$^{-1}$\,pc$^2$)$^{-1}$, vertical bar in Figure~\ref{fig:COsize_comp}).

\begin{figure}
\centering
\includegraphics[width=0.49\columnwidth]{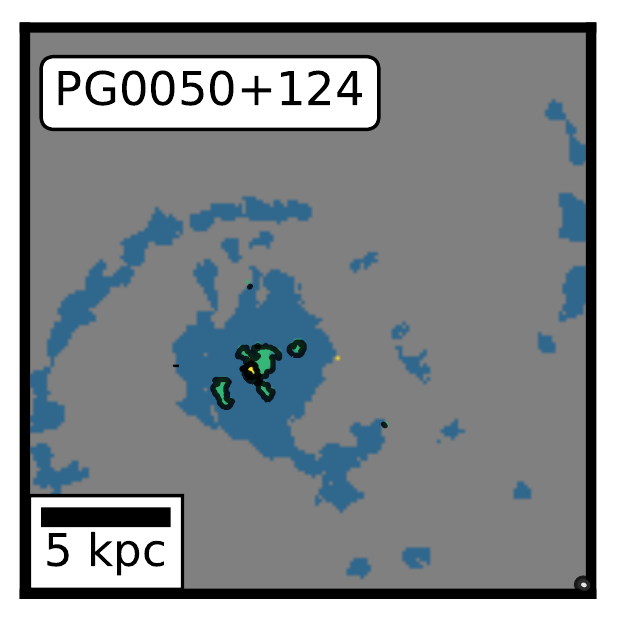}
\includegraphics[width=0.49\columnwidth]{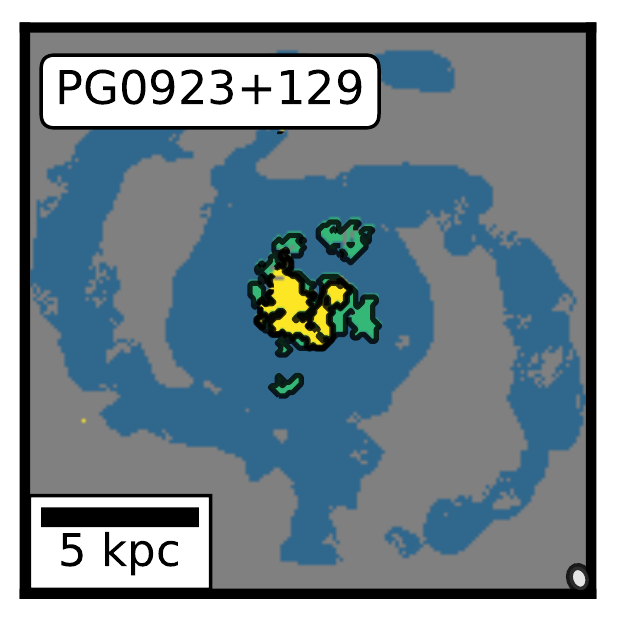}\\
\includegraphics[width=0.49\columnwidth]{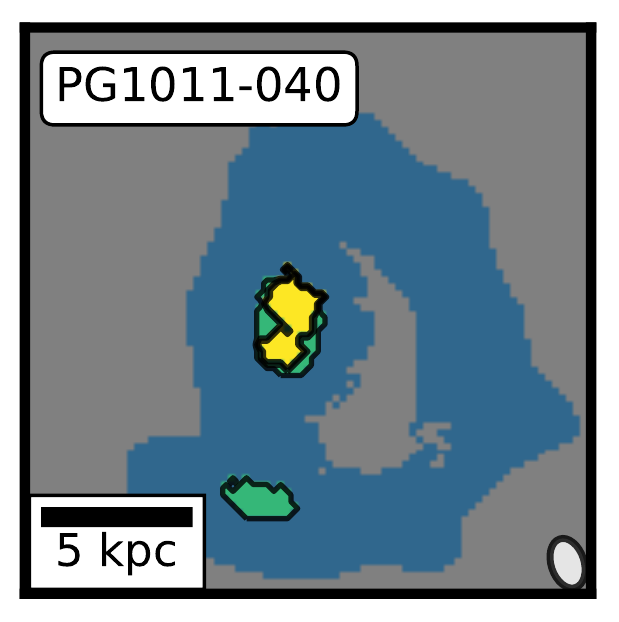}
\includegraphics[width=0.49\columnwidth]{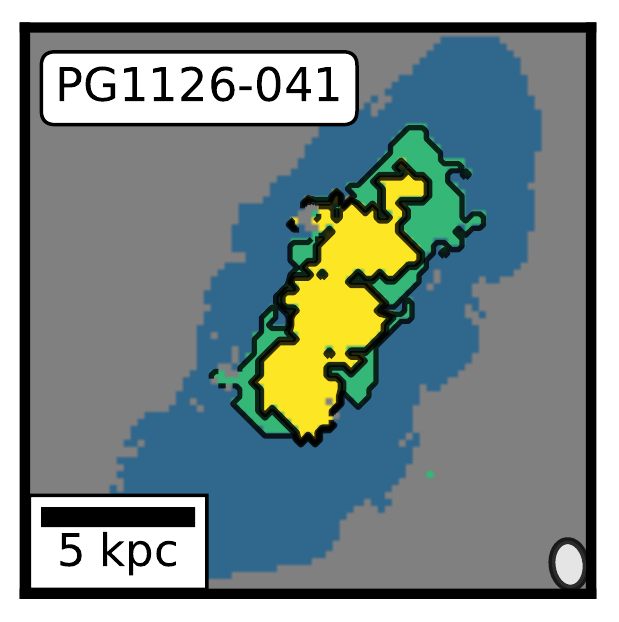}\\
\includegraphics[width=0.49\columnwidth]{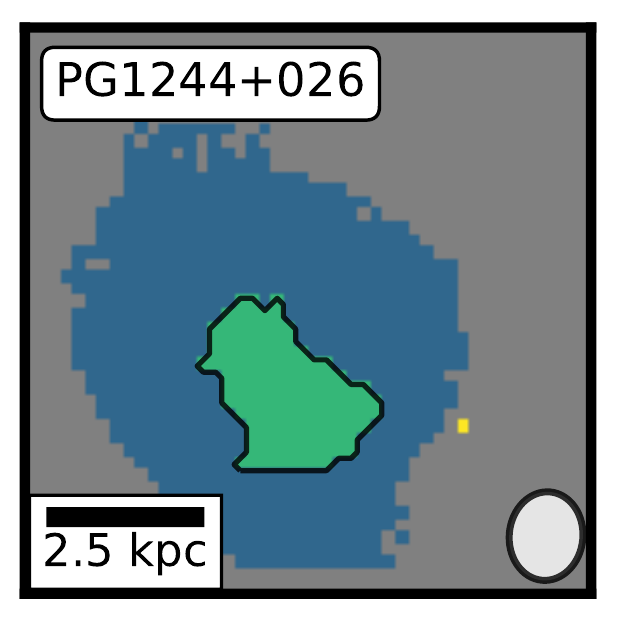}
\includegraphics[width=0.49\columnwidth]{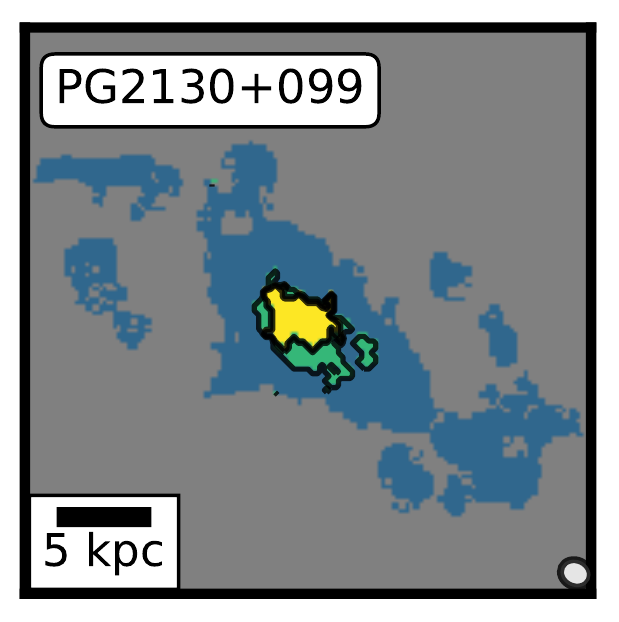}\\
\caption{\label{fig:asymmmaps} Maps of the locations of spectra with shapes classified as ``symmetric'' (dark blue), ``asymmetric'' (green), and ``complex'' (yellow), according to the multi-Gaussian fitting procedure described in Section~\ref{sec:em-lines}. In each panel, the synthesized beam is shown in the bottom right corner.}
\end{figure}

\begin{figure}
\centering
\includegraphics[width=1.0\columnwidth]{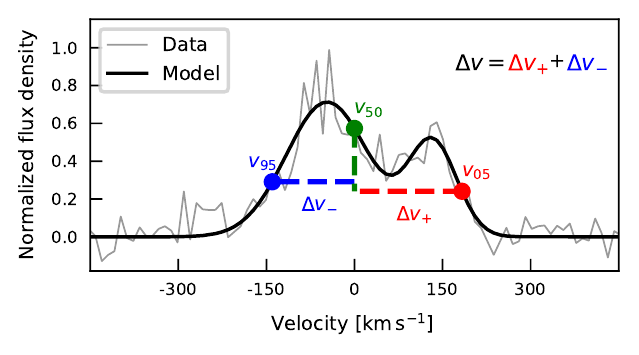}
\caption{\label{fig:emline_nonpara} Example of the non-parametric approach implemented to characterize the emission line. The quantities $v_{05}$, $v_{50}$, and $v_{95}$ correspond to the velocity values of the 5th, 50th, and 95th percentiles estimated from the cumulative distribution of the emission-line flux density. The quantities $\Delta\,v_{+}$ and $\Delta\,v_{-}$ correspond to the velocity difference of $v_{05}$ and $v_{95}$ with respect to $v_{50}$, respectively. Thus, $\Delta\,v_{+}$ is positive and $\Delta\,v_{-}$ is negative. A symmetric line is characterized by $\Delta\,v = 0$.}
\end{figure}

\begin{figure*}
\centering
\includegraphics[width=1.05\columnwidth]{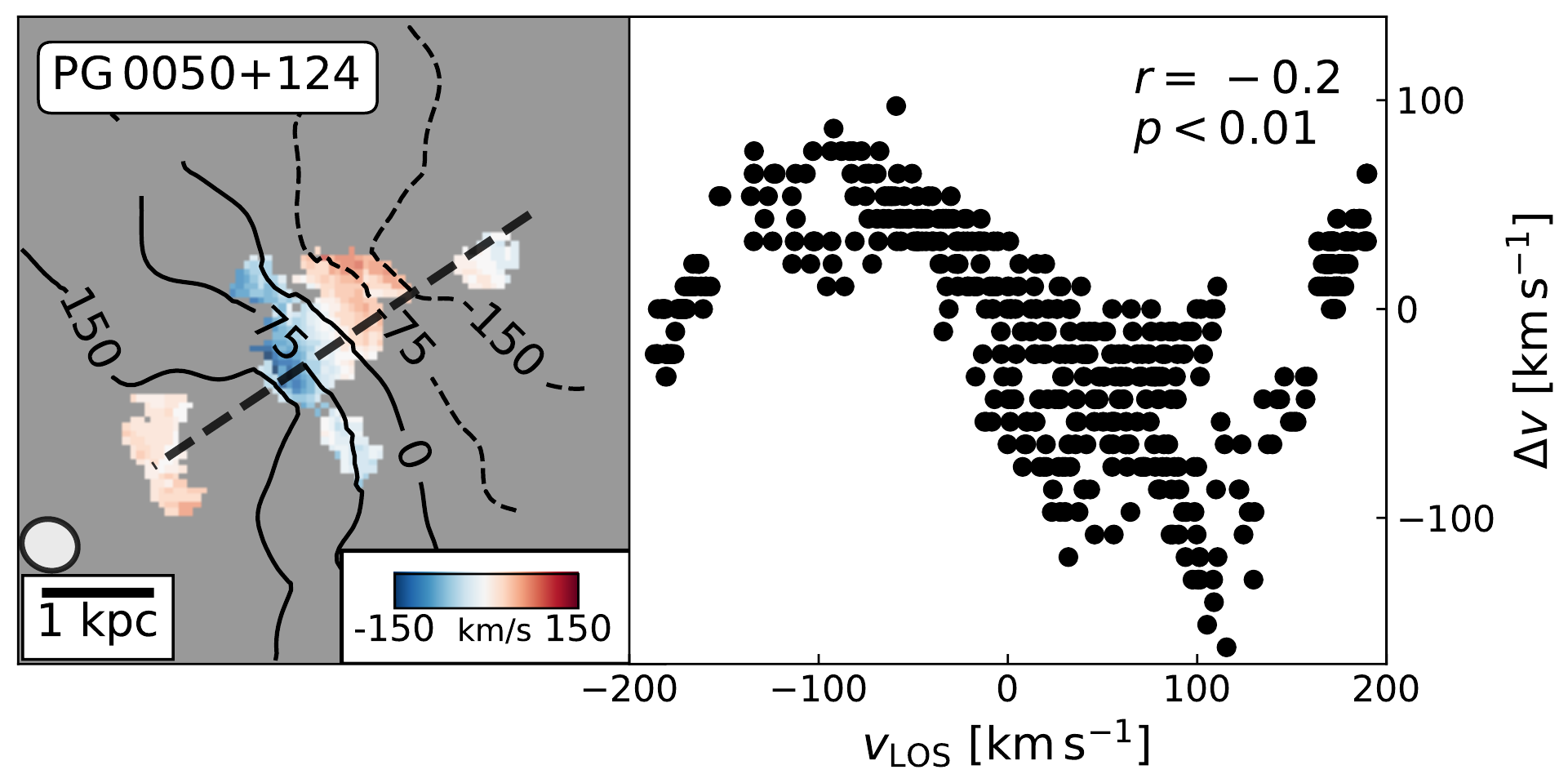}
\includegraphics[width=1.05\columnwidth]{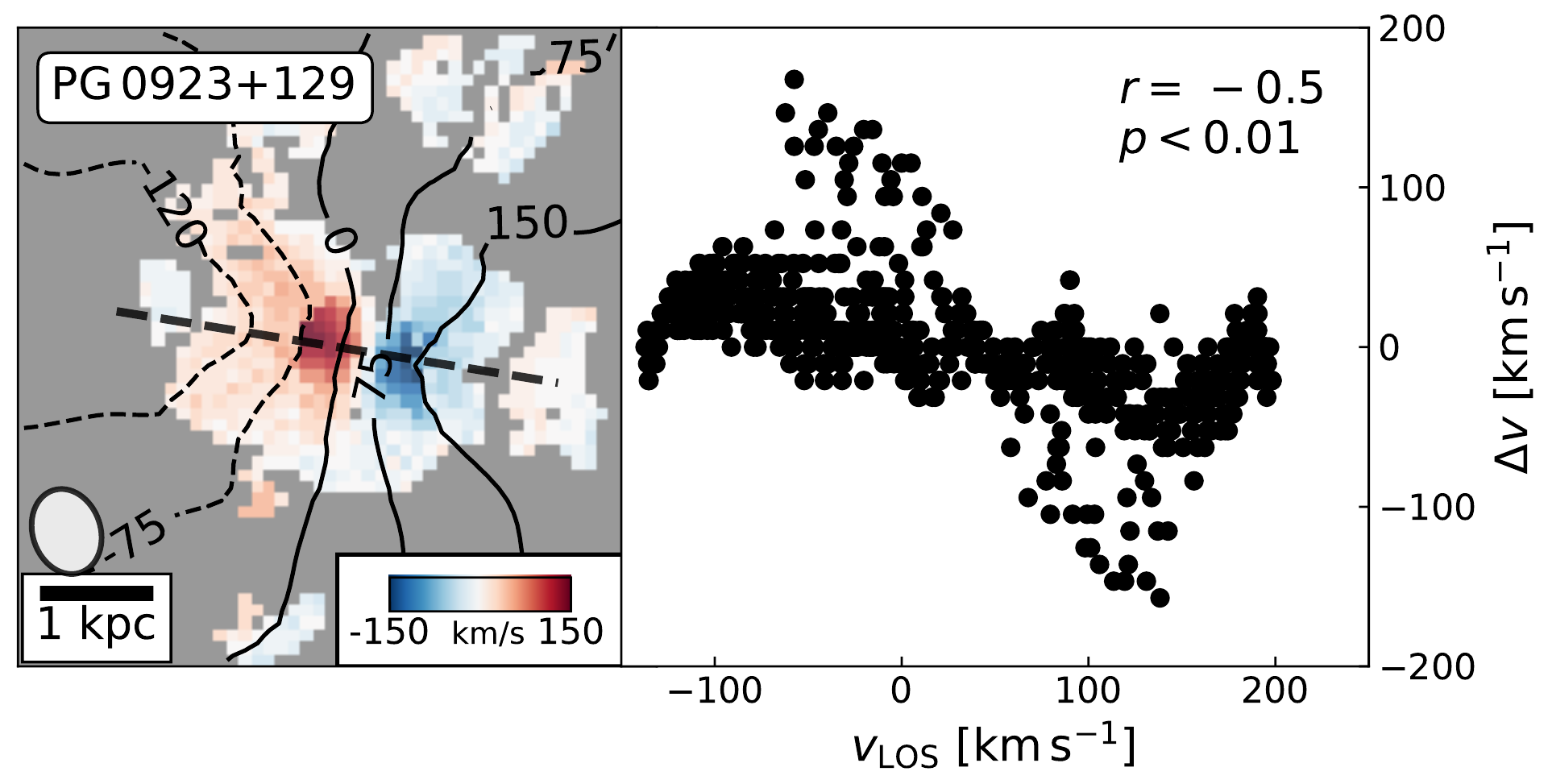}\\
\includegraphics[width=1.05\columnwidth]{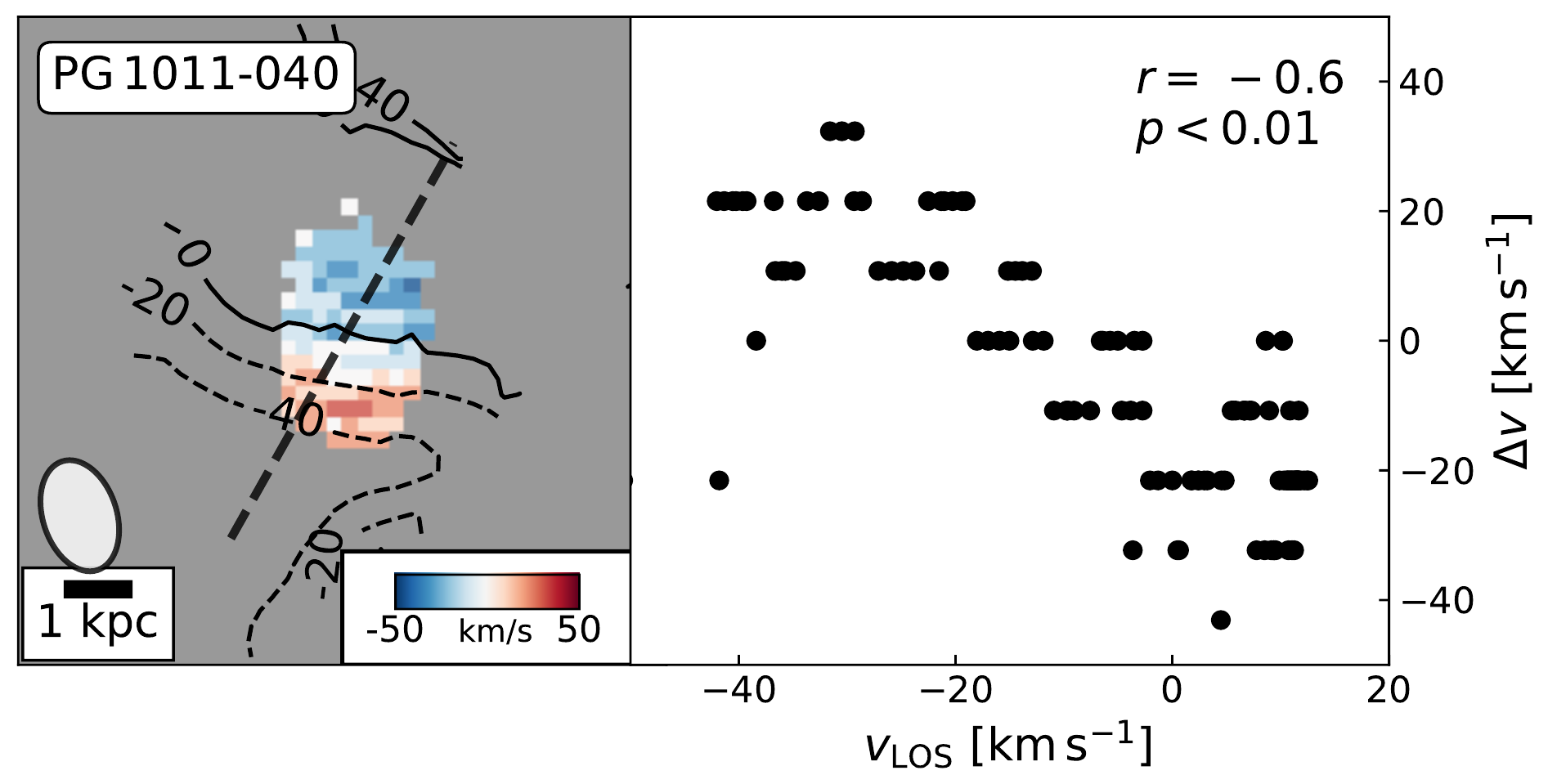}
\includegraphics[width=1.05\columnwidth]{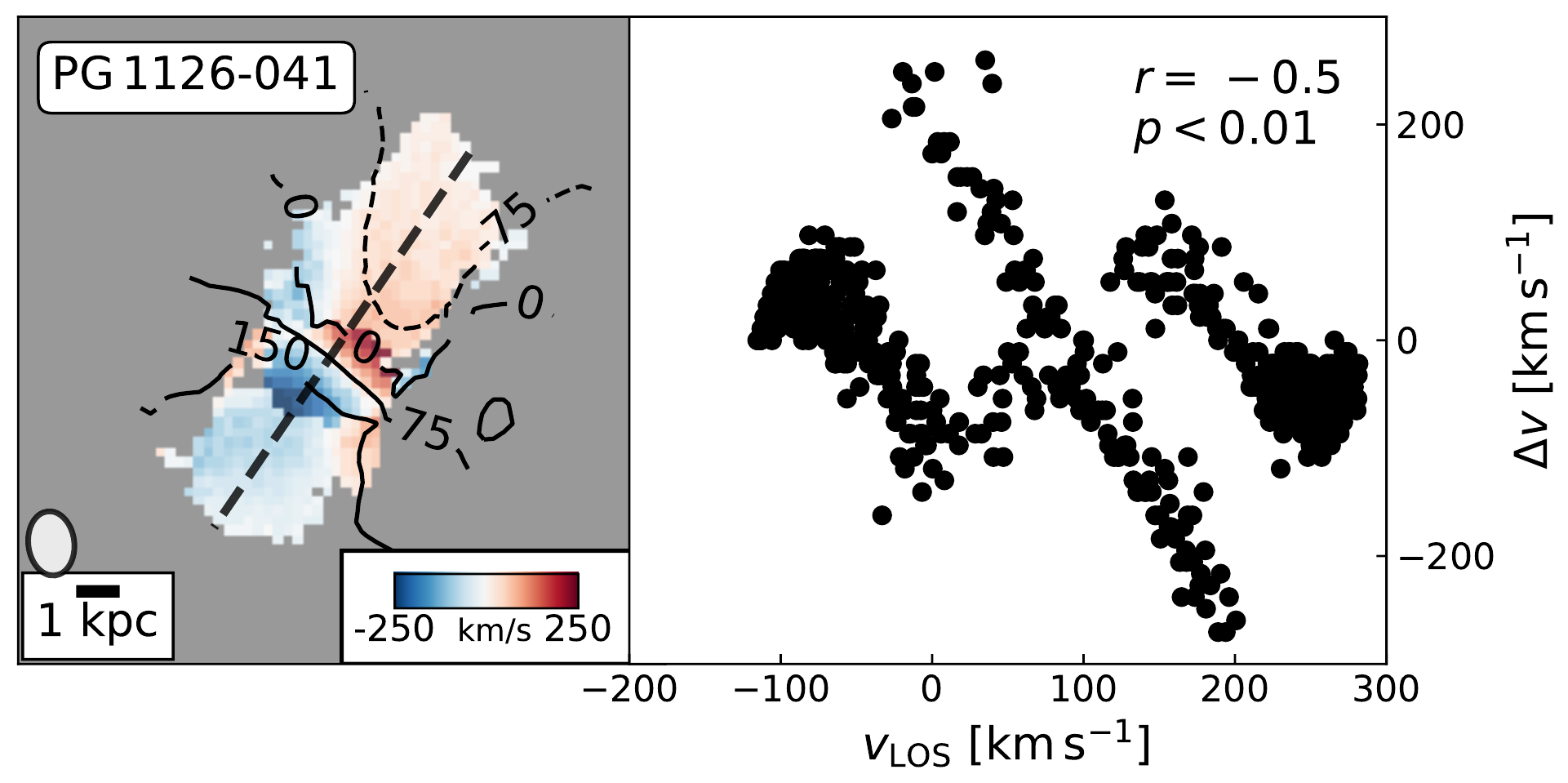}\\
\includegraphics[width=1.05\columnwidth]{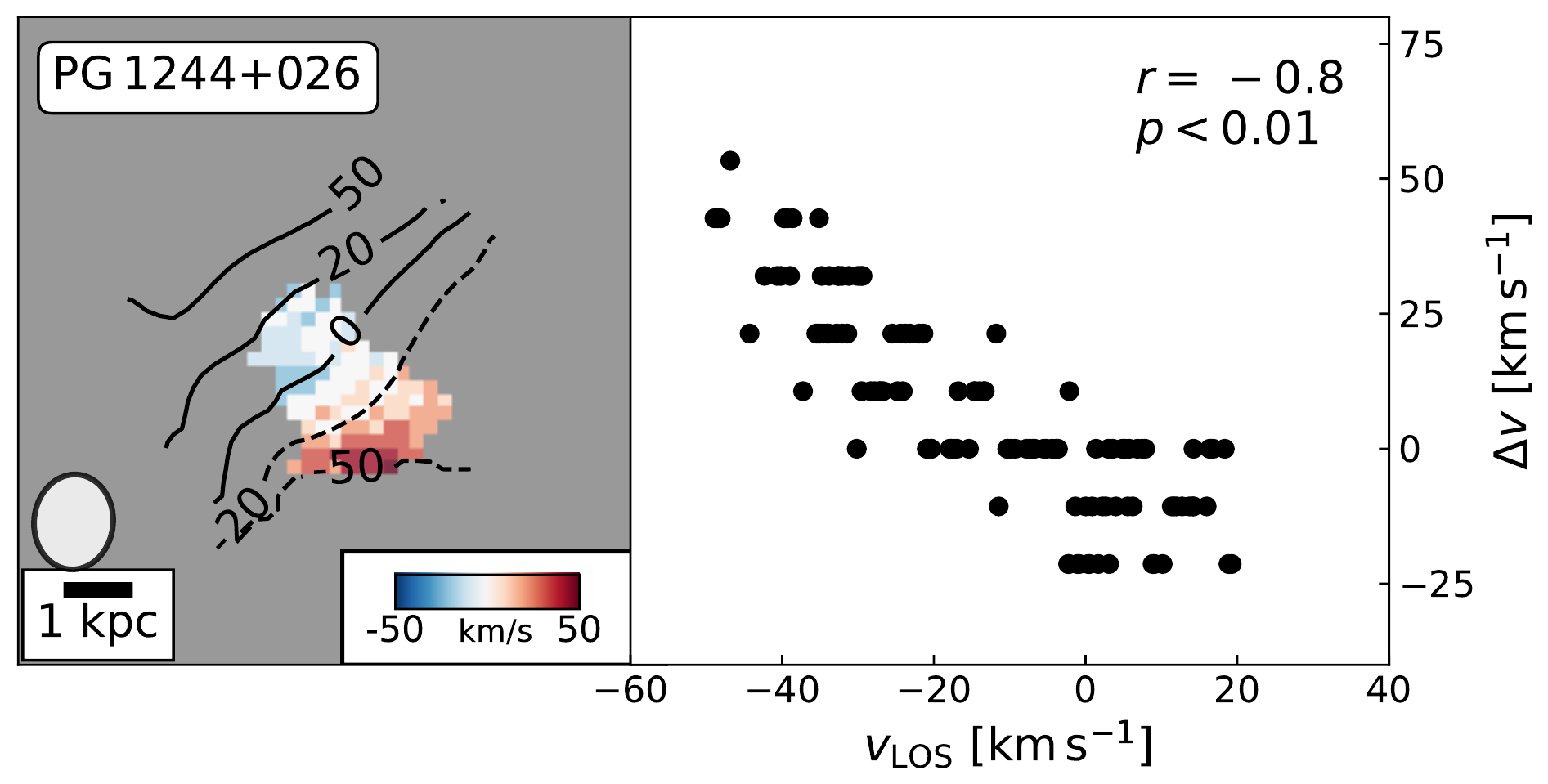}
\includegraphics[width=1.05\columnwidth]{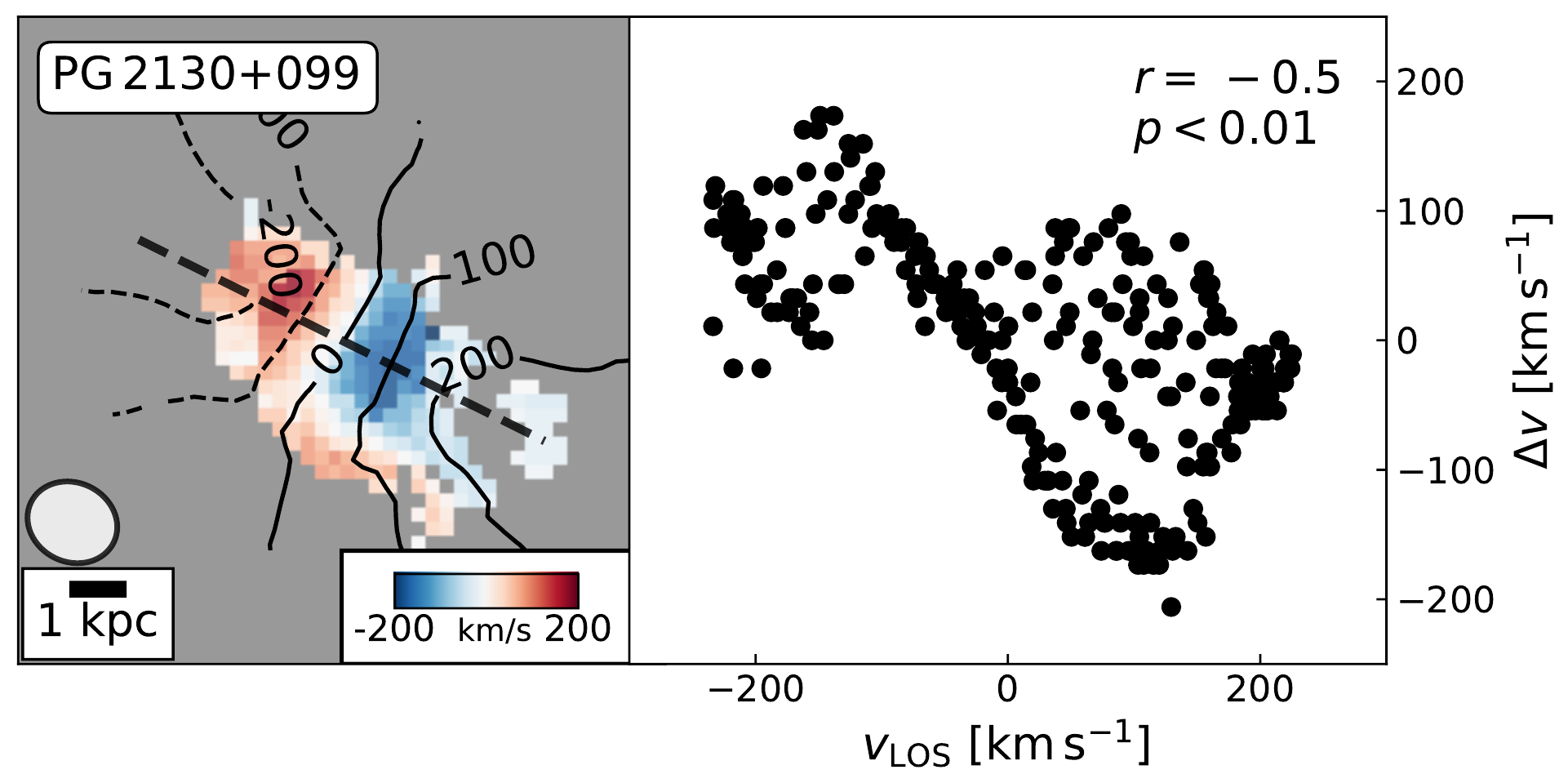}\\
\caption{\label{fig:dv_maps} Asymmetry maps of the CO(2--1) emission line for the PG quasars. The iso-velocity contours are taken from the LOS velocity maps and labeled in units of km\,s$^{-1}$, consistent with colorbar. The thick dashed line represents the kinematic major axis, when available. The synthesized beam is shown above the scale bar. To the right of each map, we show the corresponding pixel-by-pixel line asymmetry ($\Delta v$) plotted against the LOS velocity ($v_{\rm LOS}$); $\Delta v$ inversely correlates with $v_{\rm LOS}$.  We give Pearson's correlation coefficient $r$ and the $p$-value in the top-right corner. } 
\end{figure*}

\subsection{Asymmetric and Complex Line Shapes}
\label{sec:non-param}

As discussed in Section~\ref{sec:em-lines}, the pixelwise emission lines come in three generic shapes that can be decomposed into one, two, or three Gaussians, which we call ``symmetric,'' ``asymmetric,'' and ``complex,'' respectively. Figure~\ref{fig:asymmmaps} gives the spatial distribution of the line shapes, constructed from the best-fit models of the emission line in each pixel.  Symmetric lines are generally located at large radii, while complex profiles are concentrated toward the central regions of the galaxies, often associated with sub-structures in the gas distribution.

To investigate the origin of the asymmetric and complex profiles more quantitatively, we implement a non-parametric scheme to quantify the line profile, motivated by procedures commonly used to study the signatures of ionized gas outflows in galaxy spectra (e.g., \citealt{Liu2013,McElroy2015,Harrison2016,Sun2017,Husemann2019}).  To characterize the line asymmetry, we first measure the velocities at which 5\% ($v_{05}$), 50\% ($v_{50}$) and 95\% ($v_{95}$) of the total line flux is enclosed. We compute the velocity differences $\Delta v_{+} = v_{05} - v_{50}$ and $\Delta v_{-} = v_{95} - v_{50}$, which describe the line flux distribution on each side of $v_{50}$. Note that, by construction, $\Delta v_{+}$ is always positive and $\Delta v_{-}$ is always negative (Figure~\ref{fig:emline_nonpara}). Then, the line asymmetry is given simply by $\Delta v = \Delta v_{+} + \Delta v_{-}$.  A line is asymmetric when $\Delta v$ is negative or positive, which means that most of the line flux is, respectively, blueshifted or redshifted relative to its centroid. For a symmetric line, $\Delta v = 0$.  We calculate $\Delta v$ from the noiseless, best-fit emission-line model. 

Figure~\ref{fig:dv_maps} shows that $\Delta v$ varies smoothly with the LOS velocity, from negative to positive values across the maps and along the kinematic major axis of the system. Central $\sim$kpc-scale features are seen in PG\,0923+129, PG\,1126$-$041 and PG\,2130+099 systems. These seem to spatially correlate with the pixels that show complex emission-line profiles (Figure~\ref{fig:emline_shapes}) and may be produced by the presence of a central unresolved molecular gas sub-structure. In all cases $\Delta v$ correlates inversely with $v_{\rm LOS}$, with Pearson's coefficient $r = -0.2$ to $-0.8$ and very low null-hypothesis probability ($p < 0.01$).  This strongly suggests that the emission-line asymmetry is related to projected rotational motions from beam-smearing.  The observed line shape corresponds to the flux-weighted convolution of many intrinsically narrow emission lines that emanate from a projected beam-sized area on the sky, each line Doppler-shifted due to projected galactic rotation. However, due to the surface brightness distribution of the line-emitting gas, the brighter lines lie near the spectral channel that is representative of the galaxy systemic velocity (i.e., the one that is located toward the opposite side with respect to the direction of Doppler shift). Therefore, the distribution of the intrinsic emission lines in the spectrum is not symmetric, and this asymmetry is correlated inversely with the projected rotation velocity of the galaxy (see Appendix~\ref{sec:AppC} for an example).

\section{Discussion}
\label{sec:Discussion}

Although our sample is admittedly small and possibly biased (Figure~\ref{fig:PG_sample}), it nevertheless offers a first glimpse into the properties of the molecular gas in the host galaxies of nearby AGNs luminous enough to qualify as quasars.  Earlier studies have already established that, as a group, the optically/UV-selected low-redshift ($z < 0.5$) PG quasars are characteristically gas-rich \citep{Shangguan2018,Shangguan2020} galaxies forming stars with an efficiency equal to or even exceeding that of star-forming galaxies on the main sequence \citep{Shangguan2020b,Xie2020}.  What is responsible for the apparent coeval episodes of vigorous BH accretion and star formation activity?  The external environment provides no obvious clue.  While the stellar morphologies of some host galaxies suggest that they have experienced tidal interactions or recent merger activity, not all hosts with enhanced star formation show evidence of dynamical perturbations \citep{Shangguan2020b,Xie2020}.  The current ALMA sample exemplifies the problem well.  Of the three quasars that clearly lie above the $1\,\sigma$ scatter of the star-forming galaxy main sequence (see top panel of Figure~\ref{fig:comparison_sample}), only PG\,0050+124 (I\,Zw\,1) is known to belong to a multiple system \citep{Lim1999,Scharwachter2003}.  The morphologies of PG\,1126$-$041 and PG\,2130+099 resemble those of regular systems. Their CO surface brightness distribution closely follow an exponential disk distribution as the best-fit S\'ersic indexes suggest. Their velocity fields (third column of Figure~\ref{fig:int_maps}) show little evidence of perturbation (Figure~\ref{fig:kinemetry}), further confirming the regular disk-like kinematics.

The present high-resolution ALMA observations furnish some potentially useful insights.  First and foremost, we find that most of the gas resides in a compact, central disky (rotationally dominated) structure with half-light radii $\sim$1.8\,kpc.  Molecular disks of such small dimensions can be found among some star-forming galaxies in the EDGE-CALIFA survey---particularly those belonging to multiple systems \citep{Bolatto2017}---but they are not common.  The compactness of the disks, of course, naturally lead to large molecular gas surface densities, and presumably to elevated star formation activity. Such conditions are reminiscent of the conditions found in luminous and ultraluminous infrared galaxies (CO emission extension $\sim 0.3-3$\,kpc, $\Sigma_{\rm H_2} \gtrsim 300$\,$M_\odot$\,pc$^{-2}$; \citealt{Downes1998,Iono2009,Wilson2019}), as well as in the hosts galaxies of infrared-bright quasars (CO emission extension $\sim 1-7$\,kpc, \citealt{Tan2019}).  What is unusual is that our PG quasars were {\it not}\ infrared-selected nor do our selection criteria strongly biases our sample in terms of $ L_{\rm IR}$ values (or SFRs, Figure~\ref{fig:comparison_sample}), and none is experiencing a major merger event or strong tidal interactions.

Apart from the compact dimensions and high molecular gas mass surface densities, another intriguing property of the sample is the apparent misalignment between the global kinematic and photometric axes of the gas, which appears to be significant in four out of five sources, as well as the detection of significant smooth variation of the kinematic position angle with radius (kinematic twisting, \citealt{Krajnovic2008}) in three of the sources (see also \citealt{Ramakrishnan2019}). Given the lack of obvious signs of recent merging events or tidal interactions, the origin of these kinematic features is unclear.  How common are these signatures?  Do they play a role in fueling the nucleus?  Much better statistics are required, as well as observations of more luminous AGNs and more massive BHs to extend the relevant parameter range, to ultimately test against numerical simulations (e.g., \citealt{Weinberger2017,Terrazas2020}).  Comparison with a proper control sample of inactive galaxies is also essential.

\section{Conclusions}
\label{sec:Conclusions}

We present new Cycle~6 ALMA observations tracing the CO(2--1) emission line from six nearby ($z \lesssim 0.06$) Palomar-Green quasars selected from our previous Cycle~5 ACA program \citep{Shangguan2020,Shangguan2020b}. The host galaxies have normal molecular gas content ($M_{\rm H_2} \approx 10^{8.9}-10^{10.2}\,M_\odot$) for their stellar masses ($M_\star \approx 10^{10.2}-10^{11.1}\,M_\odot$) and star formation rates (SFR $\approx 2-26\, M_\odot\,{\rm yr}^{-1}$).  The ALMA observations were designed to resolve the molecular gas in the host galaxies on scales of $\lesssim 1$\,kpc (beam FWHM $\approx 0\farcs4-1\farcs4$). We concatenate the ALMA data with the previous ACA data to minimize missing flux from extended emission. Our main results are as follows:

\begin{itemize}

\item
The quasar host galaxies tend to have disk-like (S\'ersic index $n = 0.87$),  exceptionally compact (median half-light radius 1.8\,kpc) molecular gas distributions compared to inactive, star-forming galaxies of similar stellar mass, star formation rate, and molecular gas content, although some star-forming galaxies can also have similarly centrally concentrated molecular gas distributions.

\item
The velocity field of the molecular gas in five of the quasar hosts is dominated by regular rotation, with $v_{\rm rot} / \sigma_v \gtrsim 9$, and $\sigma_v = 6-36$\,\,km\,s$^{-1}$.  The remaining system (PG\,1244+026) is too compact to be analyzed with confidence. Four host galaxies exhibit a flat rotation curve out to radii $\gtrsim 8-10$ kpc from the nucleus; however, one of these systems (PG\,1011$-$040) has perturbed molecular gas kinematics.

\item
Among the five objects amenable to tilted-ring analysis, the kinematic position angle deviates from the photometric angle on average by $\sim 30^\circ$.  Three sources show evidence of a significant kinematic twist.

\item
The pixelwise spectra show a diversity of line shapes, from symmetric to asymmetric and more complex profiles.  We argue that asymmetric profiles predominantly arise from beam-smearing effects, while complex lines are predominantly associated with gas sub-structures in the central regions of the galaxies.

\end{itemize}

\acknowledgments{We thank to an anonymous referee for thoughtful comments and suggestions. We acknowledge support from the National Science Foundation of China grant 11721303 and 11991052 (LCH, RW), the National Key R\&D Program of China grant 2016YFA0400702 (LCH, RW), ANID-Chile grants Basal AFB-170002 (FEB, ET), FONDECYT Regular 1160999 (ET), 1200495 (FEB, ET) and 1190818 (ET, FEB), and Anillo de ciencia y tecnolog\'ia ACT1720033 (ET), and Millennium Science Initiative ICN12\_009 (FEB).  We thank Jing Wang for useful insights and comments. We thank MinJin Kim for sharing his stellar photometric model for PG\,2130+099 source. This paper makes use of the following ALMA data: ADS/JAO.ALMA\#2018.1.00006.S. ALMA is a partnership of ESO (representing its member states), NSF (USA) and NINS (Japan), together with NRC (Canada), MOST and ASIAA (Taiwan), and KASI (Republic of Korea), in cooperation with the Republic of Chile. The Joint ALMA Observatory is operated by ESO, AUI/NRAO and NAOJ.}

\software{\textsc{Astropy}\,\citep{astropy:2013,astropy:2018}, \textsc{CASA}\,\citep{McMullin2007}, \textsc{emcee}\,\citep{Foreman2013}, \textsc{kinemetry}\,\citep{Krajnovic2006}, \textsc{lmfit}\,\citep{Newville2014}, \textsc{matplotlib}\,\citep{Hunter2007}, \textsc{numpy}\,\citep{Oliphant2006}, \textsc{scipy}\,\citep{scipy2020}.}
\\
\appendix

\section{Notes for Individual Host Galaxies}
\label{sec:AppA}

\begin{itemize}
\item[]\textbf{PG\,0050+124:} The CO(2--1) observation shows a compact central distribution along with two extended molecular gas outer spiral arm structures that match the morphology in the restframe $I$-band HST/WFC3 image. The LOS velocity map clearly shows a disk-like rotational pattern with projected peak-to-peak rotational velocity $v_{\rm rot} \sin i \approx 440$\,km\,s$^{-1}$. The detection of the outer spiral structures allow us to trace the rotation velocities at $R \gtrsim 8$\,kpc. At this radius, the rotation curve shows lower values than in the inner zone. The velocity dispersion increases sharply at $R \approx 0.8$ and $2.8$\,kpc. We were unable to detect any CO(2--1) line emission coming from the companion systems. 

\item[]\textbf{PG\,0923+129:} The CO(2--1) line emission is distributed along a central disk-like zone from which two molecular gas spiral arms extend toward the outskirts of the HST broad-band image surface brightness distribution. We detect an inner ring-like structure. The host galaxy displays a rotational pattern, and we observe a velocity profile that extends up to the flat part of the rotation curve with projected peak-to-peak $v_{\rm rot} \sin i  \approx 330$\,km\,s$^{-1}$. The velocity dispersions increase mainly smoothly toward the galactic center, and a sharp increase is seen at $R \approx 2.5$\,kpc.

\item[]\textbf{PG\,1011$-$040:} The complex CO(2--1) morphology can be roughly separated into a central bulge-like component surrounded by a ring-like, clumpy structure. The LOS velocity map shows complex dynamics with a smooth variation of the velocity field across the field-of-view. This suggests that the two main structures are spatially connected. Comparison with the restframe $I$-band HST/WFC3 image reveals a possible spatial correlation between the molecular gas distribution and the stellar bar structure, suggesting molecular gas inflow toward the central zone due to gravitational torques.

\item[]\textbf{PG\,1126$-$041:} A clear disk-like morphology and rotational pattern can be seen. The central CO(2--1) emission seems to be distributed in a bar-like structure. The system also shows clumpy morphology. The intensity map is spatially correlated with the HST/NICMOS image, suggesting that the stellar and molecular gas components follow similar distribution. The CO emission traces the flat part of the rotation curve, from which we measure a projected peak-to-peak rotational velocity of $v_{\rm rot} \sin i  \approx 390$\,km\,s$^{-1}$. The velocity dispersion increases smoothly toward the galactic center, reaching $\sim 160$\,km\,s$^{-1}$.

\item[]\textbf{PG\,1244+026:} The CO(2--1) observation shows a compact molecular gas distribution with iso-velocity contours that depart somewhat from pure rotation. The velocity dispersion map shows very high values across most of the disk, suggesting strong beam smearing due to the compactness of the gas distribution.

\item[]\textbf{PG\,2130+099:} The regular gas distribution shows an outer spiral-like structure that matches the morphology seen in the HST/WFC3 broad-band image. Disk-like rotation can be seen from the LOS velocity map, which is sufficiently extended to allow the detection of the flat part of the rotation curve. We measure a projected peak-to-peak rotational velocity of $v_{\rm rot} \sin i  \approx 470$\,km\,s$^{-1}$ from the galaxy outskirts. The outer sub-structure seen at $R \gtrsim 10$\,kpc presents systematically lower velocity dispersion compared to the central, more disky region.

\end{itemize}

\section{Comparison of Radial Sizes}
\label{sec:AppB}

\begin{figure}
\centering
\includegraphics[width=0.9\columnwidth]{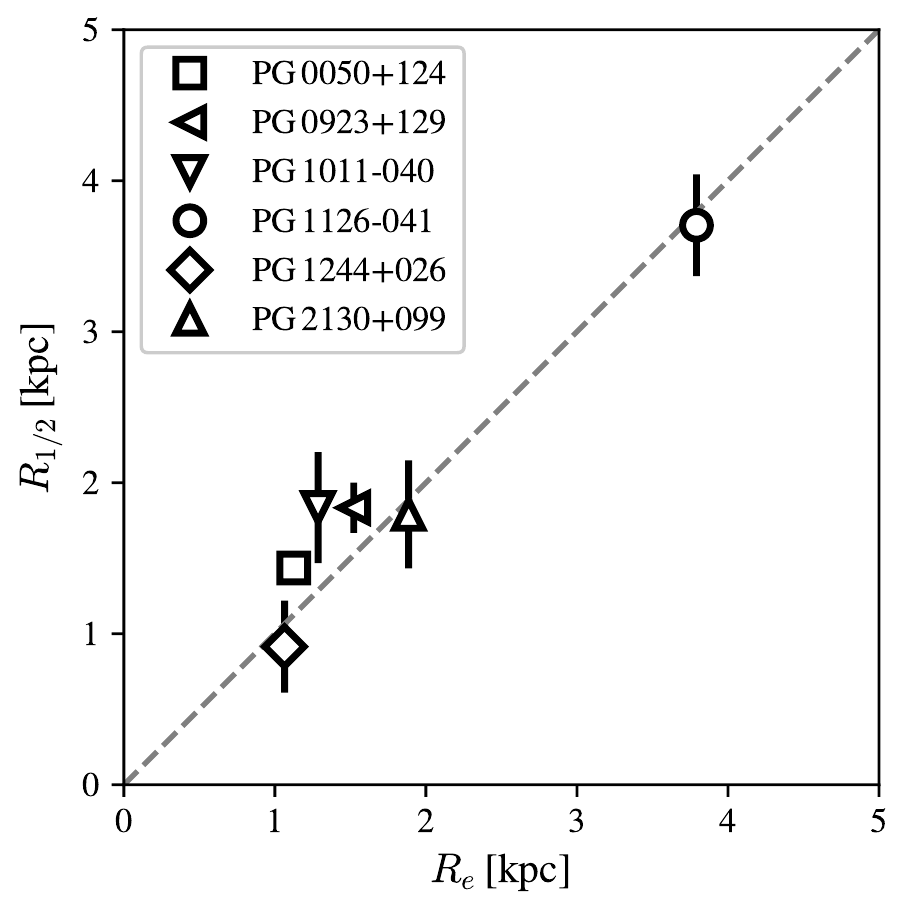}
\caption{\label{fig:Re12_Re_comp} Comparison between the half-light radius derived from the tilted-ring approach and the effective radius obtained from the best-fit S\'ersic model. We find a good agreement between both estimates.}
\end{figure}

Figure~\ref{fig:Re12_Re_comp} compares the half-light radius obtained from the best-fit S\'ersic model (i.e, the ``effective radius'' $R_e$) and the estimates obtained using the tilted-ring approach ($R_{1/2}$) following \citet{Bolatto2017}. We find a good agreement between both measurements for the six host galaxies. This suggests that the main CO(2--1) line emission components closely resemble those expected from the ideal axisymmetric two-dimensional S\'ersic profiles.

\section{Beam smearing effect on the observed emission line shapes}
\label{sec:AppC}

We simulate a galaxy observation datacube to exemplify the effect of beam-smearing on the observed emission line shapes. The simulated galaxy has an exponential surface density profile with no dark mater halo contribution. The half-light radius is set to 1.5\,kpc following our estimates reported in Table~\ref{tab:bestpar} for the host galaxies.  We assume intrinsic Gaussian emission line shapes with line widths equal to 20\,km\,s$^{-1}$ (Table~\ref{tab:kinpar}) for simplicity.  For sky-projection effects, we consider a redshift of 0.06, $b/a = 0.6$ and galaxy major axis direction parallel to the datacube $x$-axis.  Following our observational setup, we construct the datacube by setting the channel resolution equal to 11\,km\,s$^{-1}$ and the pixel size to $0\farcs2$ (Section~\ref{sec:obs}). We set an elliptical beam with FWHM$_x \times$\,FWHM$_y = 1'' \times 0\farcs1$. We choose this beam setup to simulate the beam-smearing effect only along the galaxy major axis direction and to improve the visualization of the intrinsic emission lines contribution to the resultant asymmetric emission line shape.

\begin{figure}
\centering
\includegraphics[width=1.0\columnwidth]{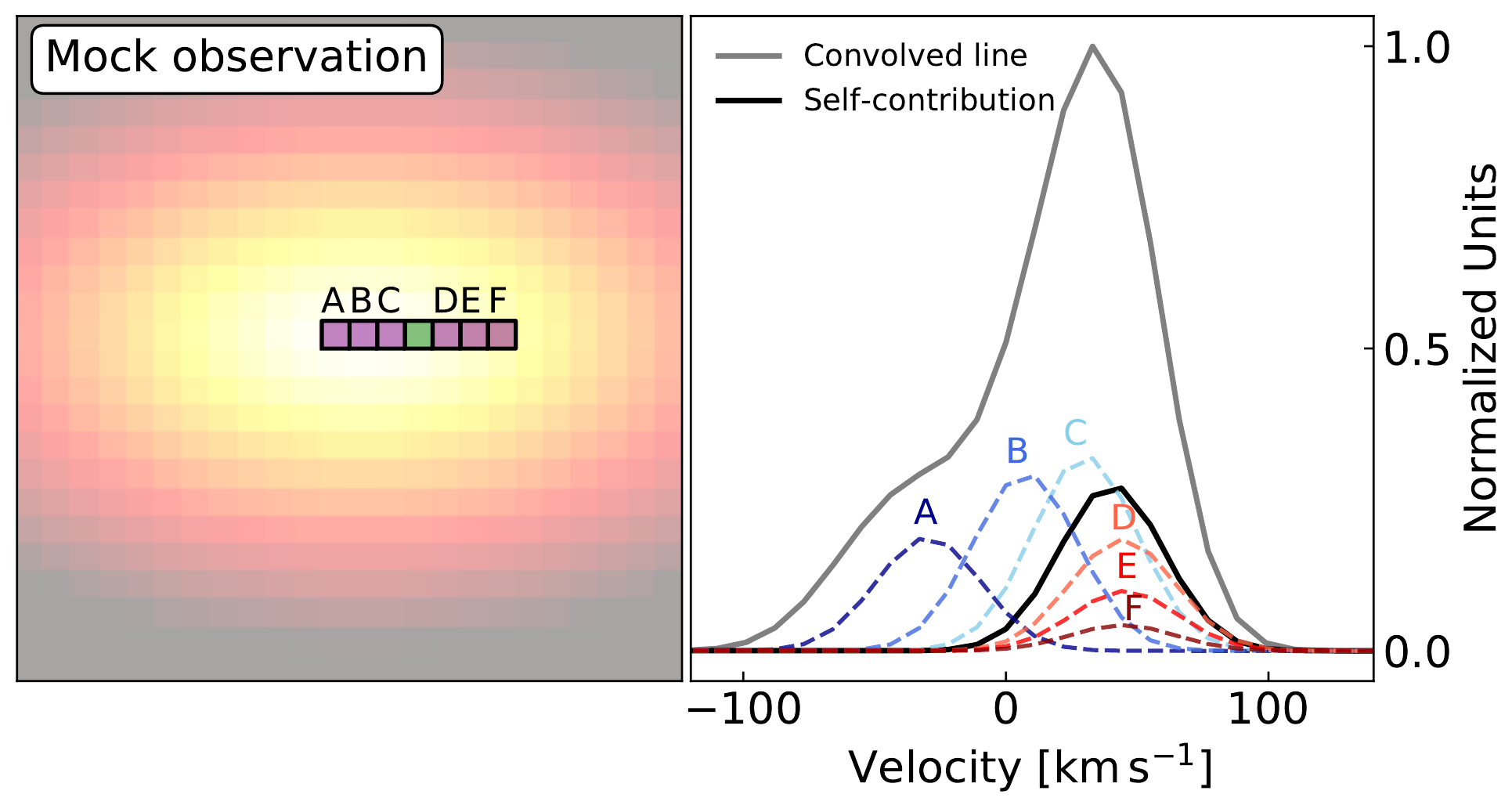}
\caption{\label{fig:Beam-Smearing_example} Example of the beam-smearing effect on the observed emission line shapes. \textit{Left:} Map of the mock galaxy observation. The green square indicates the pixel from which we extract the convolved emission line profile. The purple squares (listed by letters) correspond to the neighbor pixels that mainly contribute to the asymmetric line shape. \textit{Right:} Asymmetric emission line profile. We show the relative contribution of the intrinsic (Gaussian) emission lines measured from the target pixel (`self-contribution') and the neighbor pixels (colored-dashed lines) produced by beam-smearing.} 
\end{figure}

Figure~\ref{fig:Beam-Smearing_example} shows our simulated galaxy observation and the emission line shape extracted from the green-colored pixel. The luminosity-weighted convolution produces the emission line asymmetry as the relative contribution of the brighter neighbor pixels (representative of the systemic velocity) is greater than that of the fainter ones.  The convolved emission line tail is observed at the opposite side of the Doppler shift direction in the spectral domain.

\bibliography{example3}
\bibliographystyle{aasjournal}
\end{document}